\documentclass[10pt,aps,prb,reprint,superscriptaddress,floatfix,longbibliography]{revtex4-2}

\usepackage[T1]{fontenc}
\usepackage{physics}
\usepackage{amsmath}
\usepackage{amssymb}
\usepackage{amsthm}
\usepackage[dvipsnames]{xcolor}

\usepackage{amsbsy}
\usepackage{amstext}
\usepackage{graphicx}

\usepackage{color}
\usepackage{mathtools}
\usepackage[caption=false]{subfig}
\usepackage{multirow}

\usepackage{amsfonts}
\usepackage{dcolumn}
\usepackage{bbold,bm}
\usepackage[bookmarks=false,linkcolor=Orange,urlcolor=MidnightBlue,colorlinks,citecolor=Maroon]{hyperref}

\makeatletter
\newcommand{\hh}{\mathcal H}
\newcommand{\kv}{\vb*{k}}
\newcommand{\s}{\sigma}
\newcommand{\A}{\alpha}
\newcommand{\B}{\beta}
\newcommand{\G}{\gamma}

\newcommand{\Bt}{\widetilde{B}}
\newcommand{\id}{\mathbb{I}}
\newcommand{\cs}[1]{\cos \frac{#1}{2}}
\newcommand{\sn}[1]{\sin \frac{#1}{2}}

\newcommand\scalemath[2]{\scalebox{#1}{\mbox{\ensuremath{\displaystyle #2}}}}

\makeatother

\begin{document}
\allowdisplaybreaks

\title{Light-induced pseudo-magnetic fields in three-dimensional topological semimetals}

\author{Arpit Raj}
\email{raj.a@northeastern.edu}
\affiliation{Department of Physics, Northeastern University, Boston, MA 02115, USA}

\author{Swati Chaudhary}
\affiliation{Institute for Solid State Physics, The University of Tokyo, Chiba 277-8581, Japan}

\author{Martin Rodriguez-Vega}
\affiliation{American Physical Society, 1 Physics Ellipse Dr, College Park, MD 20740, USA}

\author{Maia G. Vergniory}
\affiliation{Donostia International Physics Center, Paseo Manuel de Lardizabal 4, 20018 San Sebastian, Spain}
\affiliation{D\'epartement de Physique et Institut Quantique, Universit\'e de Sherbrooke, Sherbrooke, J1K 2R1 Qu\'ebec, Canada}
\affiliation{Regroupement Qu\'eb\'ecois sur les Mat\'eriaux de Pointe (RQMP), Quebec H3T 3J7, Canada}

\author{Roni Ilan}
\affiliation{Raymond and Beverly Sackler School of Physics and Astronomy, Tel-Aviv University, Tel-Aviv 6997801, Israel}
\affiliation{Simons Emmy Noether Fellow, Perimeter Institute for Theoretical Physics, 31 Caroline Street North, Waterloo, ON N2L 2Y5, Canada}

\author{Gregory A. Fiete}
\affiliation{Department of Physics, Northeastern University, Boston, MA 02115, USA}
\affiliation{Quantum Materials and Sensing Institute, Northeastern University, Burlington, MA, 01803 USA}
\affiliation{Department of Physics, Massachusetts Institute of Technology, Cambridge, MA 02139, USA}
\affiliation{Department of Physics, Harvard University, Cambridge, MA 02138, USA}
% ------------------------------------------------------------------------------------
\begin{abstract}
In this work, we show that suitably designed spatially varying linearly polarized light provides a versatile route to generate and control pseudo-magnetic fields in Weyl semimetals through Floquet engineering. Within a high-frequency expansion, we derive an effective axial gauge potential $\vb{A}_5(\vb{r})$ whose curl gives the pseudo-magnetic field $\vb{B}_5(\vb{r})$. By mapping the light profile to $\vb{A}_5(\vb{r})$, we establish design principles for pseudo-magnetic field textures that mimic strain-induced gauge fields while offering key advantages like dynamic control, full reversibility, spatial selectivity, and absence of material deformation. We compare the Landau-level spectra produced by uniform real and pseudo-magnetic fields and also analyze both their linear optical conductivity and the second-order dc responses. Our results enable real-time manipulation of pseudo-magnetic fields and predict clear experimental signatures for optically engineered gauge fields in topological semimetals.
\end{abstract}
\maketitle

% ------------------------------------------------------------------------------------
\section{Introduction}

Three-dimensional topological semimetals have emerged as fascinating platforms for exploring exotic quantum phenomena, where electrons behave as relativistic fermions and exhibit linear energy dispersions near isolated band crossings~\cite{Armitage2018}. These materials include Dirac semimetals, featuring four-fold degenerate crossings protected by crystalline symmetries, and Weyl semimetals, characterized by pairs of two-fold degenerate nodes with opposite chiralities that are topologically protected and act as monopoles of Berry curvature in momentum space. These systems are highly sensitive to external perturbations including strain and other inhomogeneities which couple to the low-energy fermions as if electromagnetic fields were applied. Unlike conventional electromagnetic fields, these emergent pseudo-gauge fields act with opposite sign on the two chiralities~\cite{ilan2020pseudo}.

The most widely studied route to pseudo-gauge fields in electronic materials is strain engineering~\cite{cortijo2015prl, guinea2010natphys,levy2010science, grushin2016inhomogeneous,pikulin2016chiral}. Elastic deformations shift Weyl (or Dirac) nodes in momentum space, generating an axial vector potential $\vb{A}_{5}(\vb{r})$ whose curl produces a pseudo‑magnetic field $\vb{B}_5(\vb{r}) = \grad_{\vb{r}} \times \vb{A}_5(\vb{r})$. When $\vb{A}_{5}$ also varies in time, a pseudo‑electric field $\vb{E}_{5}=-\partial_{t}\vb{A}_{5}$ arises. The resulting strain‑induced chiral anomaly has been proposed and analyzed for both type‑I and type‑II Weyl semimetals~\cite{pikulin2016chiral, sabsovich2020pseudo}. While strain-induced pseudofields have been clearly demonstrated in two-dimensional systems such as graphene~\cite{guinea2010natphys, levy2010science} and in thin FeSn kagome films~\cite{zhang2023}, their realization in bulk three-dimensional topological semimetals remains scarce~\cite{kamboj2019}. Beyond strain engineering, magnetic textures  offer another route to generate pseudo-gauge fields~\cite{araki2020,Rogers2024,ozawa2024chiral} in magnetic Weyl semimetals where $\vb{A}_{5}$ depends on magnetization and resulting fields can reach hundreds of Tesla~\cite{ozawa2024chiral}. However, both strain and magnetic textures are material-specific and challenging to control dynamically, motivating the search for alternative approaches.

Floquet engineering, the use of time-periodic driving to modify material properties, has been extensively studied as a route to realize and control topological semimetal phases~\cite{Oka_2019,wang2014epl, hubener2017natcomm, bucciantini2017prb, chan2016prb, Zhu2024, Huang2024}. Notably, it has also been proposed as a means to generate axial gauge fields in these systems~\cite{Ebihara2016}. This axial gauge field leads to light-induced anomalous Hall conductivity which has been recently reported in the Dirac semimetal Co$_3$Sn$_2$S$_2$~\cite{yoshikawa2025light}. We can expect that the spatial modulation of light intensity which controls $\mathbf{A}_5$ should lead to non-zero $\vb{B}_5$ fields which in turn can affect the electronic transport and optical responses of the system.

The experimental probes of pseudo‑magnetic fields remain challenging because their signatures can be obscured by those of real magnetic fields ($\vb{B}$). Platforms such as photonic and acoustic metamaterials circumvent this difficulty where chiral acoustic Landau levels (LLs) have been reported~\cite{Pan2024,li2025observation}. In electronic systems, several theoretical proposals leverage the distinct symmetry properties and coupling mechanisms of $\bf B$ and ${\bf B}_5$ fields. The photovoltaic Hall effect~\cite{oka2011all} and cubic Hall viscosity~\cite{robredo2021new} have been identified as potential discriminating probes. Several studies have shown that LLs from real magnetic fields manifest directly in optical conductivity~\cite{morimoto2009prl, ikebe2010prl, shimano2011epl, shimano2013natcomm, okamura2020natcomm, jeon2014natmater, ashby2013magneto, Staalhammar2020, yadav2023magneto, liu2023landau, yadav2023magneto}, which suggests that optical probes can offer a promising route to detect LLs generated by pseudo-magnetic fields.
Recent works propose that real magnetic fields induce giant-nonlinear optical responses in Weyl semimetals~\cite{bednik2024magnetic} and we can expect to see similar effects with $\vb{B}_5$.

In this work, we investigate the generation and manipulation of pseudo-magnetic fields in Weyl semimetals using linearly polarized light within the Floquet formalism. We demonstrate that spatially modulated laser profiles induce effective axial gauge potentials that generate pseudo-magnetic fields. We focus on a particular laser profile that generates a uniform $\vb{B}_5$ leading to a distinctive Landau level structure in the effective Floquet Hamiltonian. These light-induced pseudo-magnetic fields manifest clear experimental signatures: quantum oscillations in the linear optical conductivity and a characteristic low-energy hump in the second-order nonlinear conductivity, features that are analogous to those produced by real magnetic fields. Crucially, this implies that in non-magnetic Weyl semimetals without any external magnetic fields, these optical conductivity measurements provide an unambiguous probe for detecting and characterizing pseudo-magnetic fields. Our approach offers significant advantages over strain-induced methods: dynamic control, complete reversibility, spatial selectivity, and the ability to switch pseudo-magnetic fields on ultrafast timescales without material deformation. This opens pathways for real-time manipulation of topological properties and time-resolved studies of emergent gauge phenomena in quantum materials.

Our paper is organized as follows. In Section~\ref{sec:static_ham}, we introduce the static Hamiltonian and discuss its symmetries. Sec.~\ref{sec:floquet} outlines the high‑frequency expansion used to obtain the effective Floquet Hamiltonian and examines how circularly and linearly polarized light modify the band structure of our model. Sec.~\ref{sec:LL} compares the Landau levels produced by a real magnetic field with those generated by a pseudo‑magnetic field. We present results for linear conductivity and second-order DC conductivity in Sec.~\ref{sec:conductivity_results}. Finally, we conclude with a summary of our main results in Sec.~\ref{sec:conclusion}.

% ------------------------------------------------------------------------------------
\section{Static Hamiltonian}
\label{sec:static_ham}

\begin{figure}[th!]
    \centering
    \includegraphics[scale=0.4]{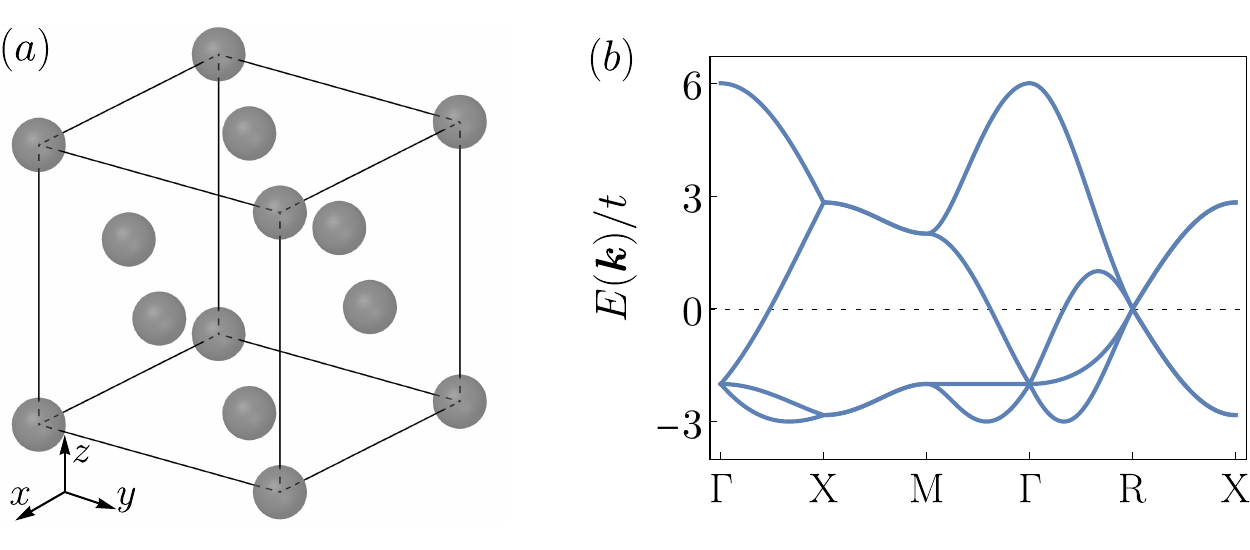}
    \caption{(a) Lattice model unit cell and (b) static energies along a high-symmetry path in the Brillouin zone.}
    \label{fig:fig1}
\end{figure}

For this study, we start by considering a material belonging to the magnetic space group \textit{P}$2_13$ with $s$-type spinless orbitals at the $4a$ Wyckoff position as introduced in Ref.~\cite{robredo2021new} but without the time-reversal symmetry (TRS) breaking magnetic flux. The tight-binding model is given in Eq.~\eqref{eq:tb-static} where $t$ is the nearest-neighbor hopping amplitude between orbitals (see Appendix~\ref{appendix:nn} for details), and $\kv = (k_x,k_y,k_z)$ is the reciprocal-space vector defined in the cubic Brillouin zone (BZ). Unless stated otherwise, each component $k_i$ should be understood as $k_ia$, with the lattice constant $a$ absorbed so that the $k$-vectors are dimensionless. As we show below, this simple model captures the essential symmetry and topology of chiral multifold semimetals in this space group and hosts a charge-2 Weyl node at the R point at energy $E=0$ that can split into two charge-1 nodes under illumination by linearly polarized light, without breaking TRS. The model also contains an opposite-chirality charge-2 Weyl node at the $\Gamma$ point but located at energy $E=-2t$, ensuring a single low-energy chiral sector near the Fermi level. Since our goal is to capture the generic features of such a system rather than fit a specific compound, we set $t=1\,\text{eV}$. In Fig.~\ref{fig:fig1}(a), we display the unit cell, and in Fig.~\ref{fig:fig1}(b) the band structure along high-symmetry lines in the BZ is shown.
\begin{widetext}
\begin{equation}
 \hh(\kv) = 2t\left[\cs{k_x}\tau_0\left(\cs{k_y}\sigma_x-\sn{k_y}\sigma_y\right)+\cs{k_y}\left(\cs{k_z}\tau_x\sigma_0-\sn{k_z}\tau_y\sigma_z\right)+\cs{k_z}\left(\cs{k_x}\tau_x\sigma_x-\sn{k_x}\tau_y\sigma_x\right)\right]
\label{eq:tb-static}
\end{equation}
\end{widetext}

Two high-symmetry points are of particular interest-- the three-fold degenerate $\Gamma$ point and the four-fold-degenerate $R$ point. Near the $R =(\pi,\pi,\pi)$ point, an expansion of the Hamiltonian to quadratic order leads to
\begin{align}
\begin{split}
\widetilde{\hh}_R(\kv) =& t \Big( \tfrac{1}{2}k_x k_y \tau_0\s_x
     + \tfrac{1}{2}k_y k_z \tau_x\s_0 + \tfrac{1}{2}k_z k_x \tau_x\s_x\\
    &\qquad + k_x \tau_0\s_y + k_y \tau_y\s_z + k_z \tau_y\s_x \Big),
\end{split}
\end{align}
with the following symmetries as determined by the software suite \textit{Qsymm}~\cite{Varjas_2018}: %identity $\id$, 
time-reversal symmetry,
two-fold rotations $R_{[100]}(\pi), R_{[010]}(\pi), R_{[001]}(\pi)$, chiral times inversion $ \mathcal{I} \mathcal{C}$, and mirror times chiral $M_{[100]}\mathcal{C}$.

The linearized Hamiltonian is given by
\begin{align}
\hh_R(\kv) &= t \big( k_x\tau_0\s_y + k_y \tau_y\s_z + k_z \tau_y\s_x \big),
\end{align}
where the symmetry $\mathcal I \mathcal C$ ($ \hh_R(\kv)  = - \hh_R(\mathcal{I} \kv)$) protects this doubly degenerate Weyl point with a total chiral charge of $-2$ (check appendix A).

Near $\Gamma$, we can write the linearized Hamiltonian as
\begin{align}
\begin{split}
    \hh_{\Gamma}(\kv) &= 2t\big( \tau_0\s_x + \tau_x\s_x + \tau_x\s_0 \big) \\ 
    &\quad - t\big(k_x \tau_y\s_x + k_y \tau_0\s_y + k_z \tau_y\s_z \big),
\end{split}
\end{align}
which possesses the cubic symmetry without inversion and with time-reversal symmetry. At the $\Gamma$ point, this leads to a non-degenerate state at energy $6t$ and three-fold degenerate states at energy $-2t$. When quadratic corrections are ignored, the Hamiltonian projected on these three states takes the form of a spin-1 Weyl system which is characterized by topological charge of +2~\cite{bradlyn2016beyond} (see Appendix~\ref{appendix:topo} for more details).

Next, we study the effect of light with different properties in the band structure. Depending on the light polarization, the degeneracies at $\Gamma$ and $R$ are lifted in different manners.

% ------------------------------------------------------------------------------------
\section{Effective Floquet Hamiltonian}
\label{sec:floquet}

We now study the effect of light with different incidence direction, polarization, frequency $\Omega$ and driving strength $A$ on the band structure of $\hh(\kv)$. We assume a classical limit, and use the minimal substitution $\kv \rightarrow \kv + \frac{ea}{\hbar}\vb{A}(t)$, where $\vb{A}(t)$ is the vector potential and $e$ is the magnitude of electron's charge. This leads to the time-dependent Hamiltonian $\hh(\kv, t) = \hh(\kv, t + 2 \pi/\Omega)$, which we treat with Floquet theory that we briefly describe below~\cite{ASENS_1883_2_12__47_0, PhysRevA.7.2203, Rodriguez_Vega_2021}.  

The remaining discrete time-translation symmetry allows us to employ the Floquet theorem~\cite{ASENS_1883_2_12__47_0} to decompose the wave function's time-dependence as $|\psi(t)\rangle= e^{i \epsilon t/\hbar}|\phi(t)\rangle$, where $|\phi(t+2 \pi / \Omega)\rangle=|\phi(t)\rangle$ and $\epsilon$ is the quasienergy. Then, the periodic piece obeys the Floquet-Schr\"odinger equation 
\begin{equation}
\left[\hh(t)-i \hbar\partial_{t}\right]|\phi(t)\rangle=\epsilon|\phi(t)\rangle.
\end{equation}

We can obtain the quasienergies and steady states from the Floquet evolution operator $U_{F}=\mathrm{\hat{T}} \exp \left\{-\frac{i}{\hbar} \int_{0}^{2 \pi / \Omega} \hh(s) d s\right\}=e^{-i \hh_{F} T/\hbar}$, where $\mathrm{\hat{T}}$ denotes time ordering, $T=2\pi/\Omega$ is the period, and $\hh_{F}$ is the effective Floquet Hamiltonian governing stroboscopic dynamics. Alternatively, we can employ the extended-state picture which follows from a Fourier expansion of the steady states  $|\phi(t)\rangle=\sum_{n} e^{i n \Omega t}\left|\phi_{n}\right\rangle$.  This leads to
\begin{equation}
    \sum_{m}\left(\hh^{(n-m)}+\delta_{n, m} \hbar\Omega m\right)\left|\phi_{m}\right\rangle=\epsilon\left|\phi_{n}\right\rangle,
\end{equation}
with Hamiltonian Fourier modes given by $\hh^{(n)}=\frac{\Omega}{2\pi}\int_{0}^{2 \pi/\Omega}\dd{t}  \hh(t) e^{-i n \Omega t}$. The latter representation is particularly useful to obtain an effective Floquet Hamiltonian in the high-frequency regime. In particular, we perform a Van Vleck approximation~\cite{Eckardt_2015, Rodriguez_Vega_2021}. In this work, we consider the effective Hamiltonian up to first order in $1/\Omega$,
\begin{align}
\hh^{F} \approx \hh^{(0)}+\sum_{m\neq 0}\frac{ {\hh}^{(m)} {\hh^{(-m)} }}{m\hbar\Omega},
\label{eq:effectiveFloquet}
\end{align}
where $\hh^{(0)}$ is the averaged Hamiltonian over one cycle. 

In the next section we derive the effective Hamiltonians near the special points $\Gamma$ and $\rm R$ for different light polarization.

% ------------------------------------------------------------------------------------
\subsection{Circularly polarized light}
\label{ssec:cpl}

First, we study the effect of circularly polarized light on the band structure. We consider the vector potential
\begin{align}
    \vb{A}(t) = A_0 [\sin (\Omega t) \vu{y} + \cos (\Omega t) \vu{z}].
    \label{eq:circularA0}
\end{align}
As a starting point, and to gain initial insight into the effect of light in this case, we obtain and analyze the linearized effective Floquet Hamiltonians near the $\Gamma$ and $\rm R$ points.

Near the $\rm R$ point, we find ($C$ in the superscript denotes circularly polarized light),
\begin{align} 
\hh^{F,C}_{R}(\kv) &= t((k_x+\tfrac{\G^2}{\hbar\Omega/t}) \tau_0\s_y  +  k_y \tau_y \s_z  +  k_z \tau_y \s_x),
\end{align}
where $\gamma=\frac{A_0 a}{\hbar/e}$. This leads to the quasienergies
\begin{equation}
E_R^{\pm} = \pm t\, \sqrt{\left(k_x+\tfrac{\G^2}{\hbar\Omega/t}\right)^2 + k_y^2 + k_z^2}.
\end{equation}
Therefore, circularly polarized light only shifts the four-fold degeneracy away from the $\rm R$ point, without opening up a gap. 

Near the $\Gamma$ point, the Hamiltonian takes the form
\begin{align}
\begin{split}
    \mathcal{H}^{F,C}_{\Gamma}(\kv) &= 2t ( \tau_0\s_x + \tau_x \s_x + \tau_x\sigma_0\\ 
    &\quad - \tfrac{1}{2}\left(k_x - \tfrac{\G^2}{\hbar\Omega/t}\right) \tau_y\s_x - \tfrac{1}{2}k_y \tau_0 \s_y - \tfrac{1}{2}k_z \tau_y \s_z ).
\end{split}
\end{align}
The quasienergies are given by
\begin{align}
   E_{\Gamma,1}^\pm &= -2t \pm t\,\sqrt{\left(k_x-\tfrac{\G^2}{\hbar\Omega/t}\right)^2 + k_y^2 + k_z^2},\\
   E_{\Gamma,2}^\pm &= 2t \pm t\,\sqrt{\left(k_x-\tfrac{\G^2}{\hbar\Omega/t}\right)^2 + k_y^2 + k_z^2 + 16},
\end{align}
which, again, only shifts the three-fold degeneracy away from the $\Gamma$ point, without opening up a gap.

\begin{figure}[t]
    \centering
    \includegraphics[scale=0.4,trim={0 0.3cm 0 0},clip]{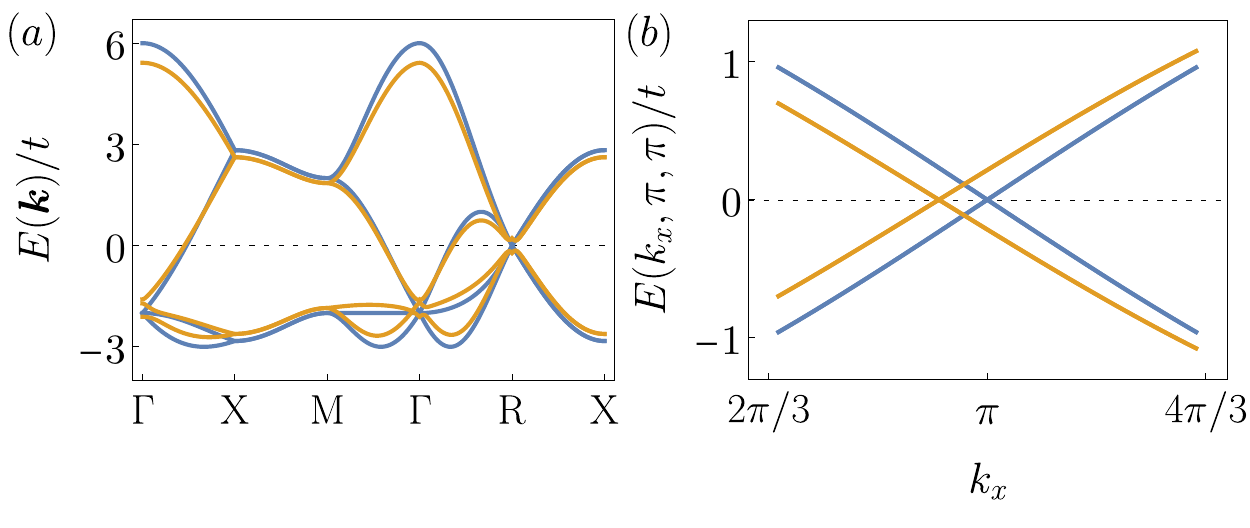}
    \caption{(a) Exact quasienergies for a model driven with circularly polarized light along a high-symmetry path for $\gamma=1.1$, $\hbar\Omega/t=5$ (orange), compared with the static energies (blue). (b) Quasienergies near the R point as a function $k_x$, which shows that the node shifts but does not split.}
    \label{fig:fig2}
\end{figure}

In the linearized models, circularly polarized light only shifts the degenerate points in the Brillouin zone. In the static system, TRS pins the two chiralities to two different time-reversal invariant momenta (TRIM). However, circularly polarized light breaks TRS and can move the degenerate nodes away from the TRIM. This kind of effect of circularly polarized light is significantly different from previously proposed schemes in other models where it split a Dirac node into two Weyl nodes~\cite{bucciantini2017prb, hubener2017natcomm}. 

Now, we compare these results with the exact quasienergies obtained using the effective Floquet Hamiltonian in the extended space (see Appendix~\ref{appendix:heffCP}). 
We show our results in Fig.~\ref{fig:fig2}. The plot along the high-symmetry path suggests that the bands at the R point are gapped out. However, a three-dimensional quasienergy plot reveals that the four-fold degeneracy is not lifted, but only shifted away from the $\rm R$ point along $k_x$, as indicated from the low-energy model.

% ------------------------------------------------------------------------------------
\subsection{Linearly polarized light}
\label{ssec:lpl}

\begin{figure}[b]
    \centering    \includegraphics[scale=0.4,trim={0 0.3cm 0 0},clip]{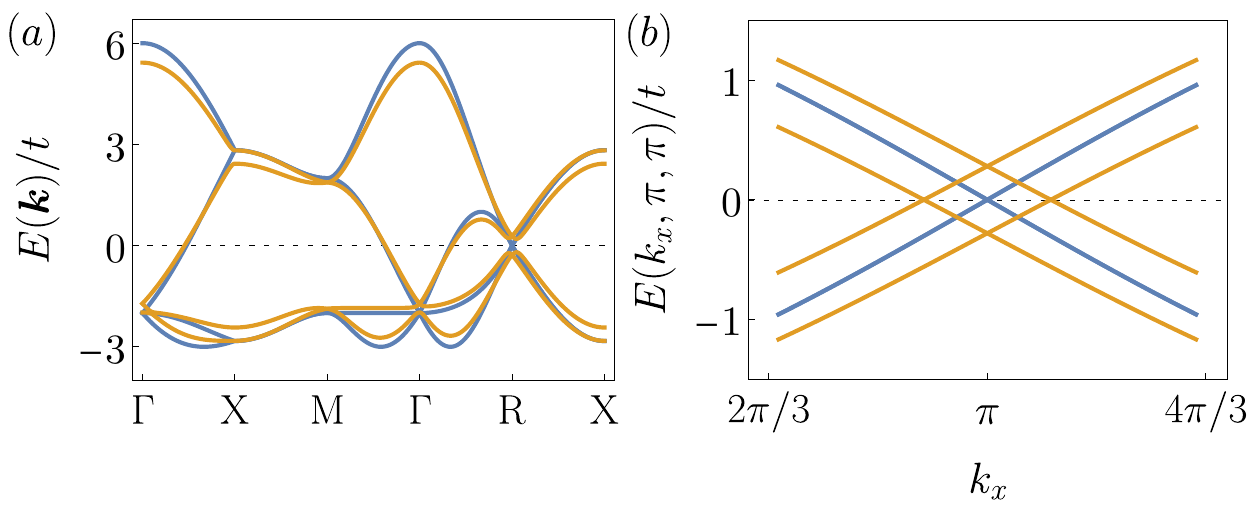}
    \caption{(a) Exact quasienergies for a model driven with linearly polarized light along a high-symmetry path for $\G=1.1$ (orange), compared with the static energies (blue). (b) Quasienergies near the R point as a function $k_x$, which shows splitting of the node. }
    \label{fig:fig3}
\end{figure}

Now, we consider the effect of linearly polarized light. We consider the vector potential
\begin{align}
    \vb{A}(t) &= A_0 \sin (\Omega t) (\vu{y}+\vu{z}),
    \label{eq:linearA0}
\end{align}
which is included in the Hamiltonian in Eq.~\eqref{eq:tb-static} via the Peierls substitution \cite{Rodriguez_Vega_2021}. The effective Floquet Hamiltonian, up to first order in $1/\Omega$, is given by (where the $L$ in the superscript denotes linearly polarized light),
\begin{widetext}
\begin{align}
\begin{split}
 \hh^{F,L}(\kv) =& t\bigg[ 2J_0\left(\frac{\G }{2}\right)\cs{k_x}\tau_0\left(\cs{k_y}\sigma_x-\sn{k_y}\sigma_y\right) + 2J_0\left(\frac{\G }{2}\right)\cs{k_z}\left(\cs{k_x}\tau_x\sigma_x-\sn{k_x}\tau_y\sigma_x\right) \\
 &\quad + \left(\cs{k_z-k_y}+J_0(\G) \cs{k_z+k_y}\right)\tau_x\sigma_0 - \left(\sn{k_z-k_y}+J_0(\G)\sn{k_z+k_y}\right)\tau_y\sigma_z \bigg ]
\end{split}
\label{eq:driven0}
\end{align}
\end{widetext}
where, $J_0(x)$ is the zeroth-order Bessel function of the first kind, and $\G=\frac{A_0a}{\hbar/e}$. We note that the $1/\Omega$ term is zero in this case. As $A_0$ is increased from zero, the node at R splits into two and the two nodes move to $(k^{\pm}_0,\pi,\pi)$ where $k^{\pm}_0=2\cos^{-1}(\pm\frac{J_0(\G)-1}{2J_0(\G/2)})=\pi\pm\frac{\G^2}{4}\pm O(\G^6)$. Here, $\pm$ refer to nodes that move to the right and left of $k_x=\pi$, respectively. Since $\pm\frac{J_0(\G_0)-1}{2J_0(\G_0/2)}=\mp 1$ at $\G_0\approx 2.746421938$, $k_0^\pm$ is only defined for $\G<\G_0$. In Fig.~\ref{fig:fig3}(a), we plot the quasienergy spectrum for $\hh^{F,L}$ with $\gamma=1.1$ along a high-symmetry path. The node splitting at R is shown in Fig.~\ref{fig:fig3}(b). We should emphasize here that our choice of $\gamma=1.1$ is made for simplicity and to highlight the essential physics. A larger $\gamma$ increases the node separation and simplifies optical-response calculations even on a coarse $k$-mesh. With $a\sim 5$~\AA, which is representative of the lattice constants of several Weyl semimetals in the space group~\textit{P}$2_13$~\cite{xu2020optical, maulana2020,eriksson2004_Mn3IrGe, eriksson2004_Mn3IrSi} and $\hbar\Omega \sim 5\,\rm eV$, the peak electric field is on the order of $1 \text{V/\AA}$ which corresponds to laser intensity  of $10^{13}\,\rm W/cm^2$, attainable with ultrafast pulsed lasers but difficult to sustain under continuous illumination. In an actual material realization, more experimentally accessible conditions could be achieved by using smaller drive strength $A_0$. For example, taking $\gamma=0.1$ with the same drive frequency would reduce the laser intensity to about $10^{11}\,\rm W/cm^2$.

Near the R point, we can also look at the low-energy effective Floquet Hamiltonian
\begin{align}
\begin{split}
\widetilde{\hh}^{F,L}_R(\kv) =& \frac{t}{2} \Big( k_x k_y  \tau_0 \s_x + 2 k_x \tau_0 \s_y + (k_y k_z+\tfrac{\G^2}{2}) \tau_x \s_0 \\
&\quad + 2 k_y \tau_y \s_z+ k_z k_x \tau_x \s_x + 2  k_z \tau_y \s_x\Big),
\end{split}
\label{eq:efh_LR}
\end{align}
where the light-induced term $\frac{t\G^2}{4} \tau_x\s_0$ originates from the time average of the quadratic term $k_y k_z \tau_x\s_0$ and breaks $ \mathcal{I} \mathcal{C}$. Considering only terms linear in momentum, we can apply a series of unitary transformations (see Appendix~\ref{appendix:transformation}) to express the effective Floquet Hamiltonian as
\begin{align} 
\hh^{F,L}_R(\kv) &= t\left( k_x \tau_0\s_y - k_y \tau_0\s_x + k_z \tau_0\s_z  + \tfrac{\G^2}{4}\tau_z \s_y\right),
\label{eq:efh_LR_lin}
\end{align}
which has a block-diagonal form, with the two blocks corresponding to Weyl Hamiltonians $(k_x\pm \tfrac{\G^2}{4})\s_y- k_y\s_x + k_z\s_z$. These describe Weyl nodes of the same chirality, symmetrically displaced from the R point along $k_x$.

\begin{figure}[t!]
    \centering
    \includegraphics[scale=0.4]{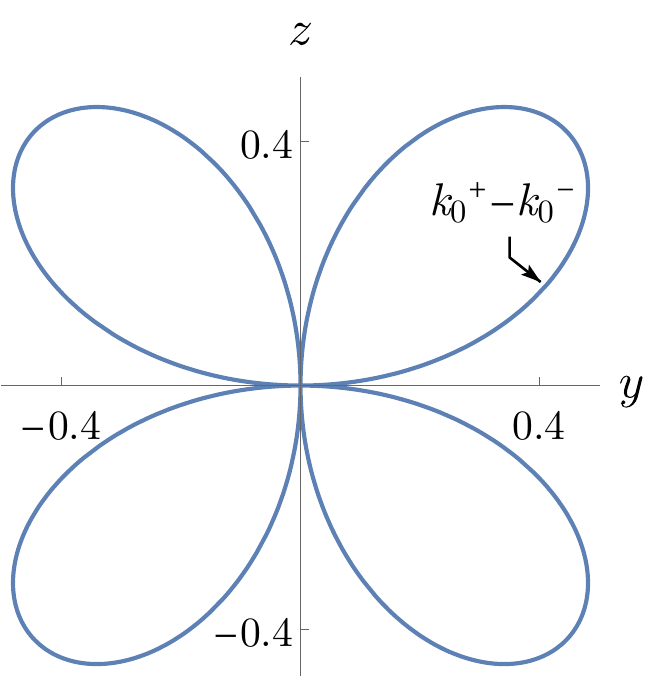}
    \caption{Polar plot of Weyl-node separation $(k_0^+-k_0^-)$ vs. polarization angle $\theta$ of light in y-z plane. We have used $\G=1.1$.}
    \label{fig:polarization}
\end{figure}

For completeness we also look at the dependence of the node separation on the polarization angle. We consider 
\begin{align}
    \vb{A}(t) &= \sqrt{2}A_0\sin(\Omega t) (\cos{\theta}\,\vu{y}+\sin{\theta}\,\vu{z}),
\end{align}
which gives Eq.~\eqref{eq:linearA0} for $\theta=\pi/4$. Using this, we obtain
\begin{align}
k^{\pm}_0=2\cos^{-1}\bigg[\pm\frac{J_0(\G\frac{\cos{\theta}+\sin{\theta}}{\sqrt{2}})-J_0(\G\frac{\cos{\theta}-\sin{\theta}}{\sqrt{2}})}{2J_0(\G\frac{\cos{\theta}}{\sqrt{2}})}\bigg].
\end{align}
The dependence of $(k_0^+-k_0^-)$ on polarization angle $\theta$ is shown in Fig.~\ref{fig:polarization}. The R-node splits into two only when the polarization is off-axis in the $yz$-plane.

The results derived in this section indicate that by using linearly polarized light we can break the chiral multifold Weyl cone's four-fold degeneracy at the $\rm R$ point and obtain two two-fold degenerate Weyl points of the same chirality. Most importantly, the $k$-space separation between these Weyl points is directly determined by the light amplitude and becomes crucial to generating a pseudo-magnetic field in the system.

% ------------------------------------------------------------------------------------
\section{Landau levels}
\label{sec:LL}

\subsection{Pseudo-magnetic field}
\label{ssec:ll_B5}

Before looking at the Landau levels, we briefly describe how the model in Eq.~\eqref{eq:driven0} gives rise to pseudo-magnetic fields. In the typical Floquet approach, one assumes that the vector potential $\vb{A}(t)$ is spatially constant. However, in experimental implementations, we can expect to obtain a position-dependent vector potential amplitude. This can arise from engineered laser beam profiles using various beam shaping techniques~\cite{dickey2014laser, laskin2013}, or the attenuation of the light inside a material. If these variations are on length scales large compared with the unit cell size, we can define a local band structure with a position-dependent vector potential.

For this study, we consider a linearly polarized light described by a position dependent vector potential
\begin{align}
    \vb{A}(y,t) = A(y) \sin (\Omega t) (\vu{y}+\vu{z}),
    \label{eq:linAy}
\end{align}
with
\begin{align}
    A(y) = \widetilde{\G}_0 \sqrt{\sin(\tfrac{\alpha+A_1 y}{2})},
    \label{eq:Ay}
\end{align}
where we have defined $\widetilde{\G}_0=\frac{\hbar}{e a}\G_0$ and $\alpha=2\sin^{-1}(\frac{A_0^2}{\widetilde{\G}_0^2})$ with $A_0<\widetilde{\G}_0$ (hence, $0\leq \A<\pi$). We take the system to have $N_y$ unit cells along the $y$-axis, with the origin situated midway, and consider open boundary conditions. Since the system is centered at $y=0$, $A(y)$ in Eq.~\eqref{eq:Ay} is well defined if $|A_1|<{\rm min}\left(\frac{2\A}{N_y a},\frac{2(\pi-\A)}{N_y a}\right)$. Note that when $A_1=0$, we get $A(y)=A_0$. 

Having a position dependent amplitude gives rise to a pseudo-magnetic field along $\vu{z}$,
\begin{align}
\vb{B}_5(\vb{r}) = \grad_{\vb{r}} \times \vb{A}_5(\vb{r}) ,
\end{align}
where 
\begin{align}
    \vb{A}_5(\vb{r}) = \frac{2\hbar}{e\,a}\,\cos^{-1}\left(\frac{J_0\left(\tfrac{aA(y)}{\hbar/e}\right)-1}{2J_0\left(\tfrac{aA(y)}{2\hbar/e}\right)}\right) \vu{x} .
\end{align}
Our main motivation behind the particular choice of $A(y)$ in Eq.~\eqref{eq:Ay} is to generate a near uniform pseudo-magnetic field for better comparison with a uniform real magnetic field. With this choice, we get $\vb{B}_5\approx -\frac{\hbar}{ea^2} \frac{\G_0^2}{8} A_1a\,\vu{z}$. For $A_1=0.004/a$, this gives $B_5\sim 10\,\rm T$ which corresponds to pseudo-magnetic length $l_{B_5}=\sqrt{\frac{\hbar}{eB_5}}\sim 10\,\rm nm$.

\begin{figure}[t]
    \centering
    \includegraphics[width=0.98\linewidth]{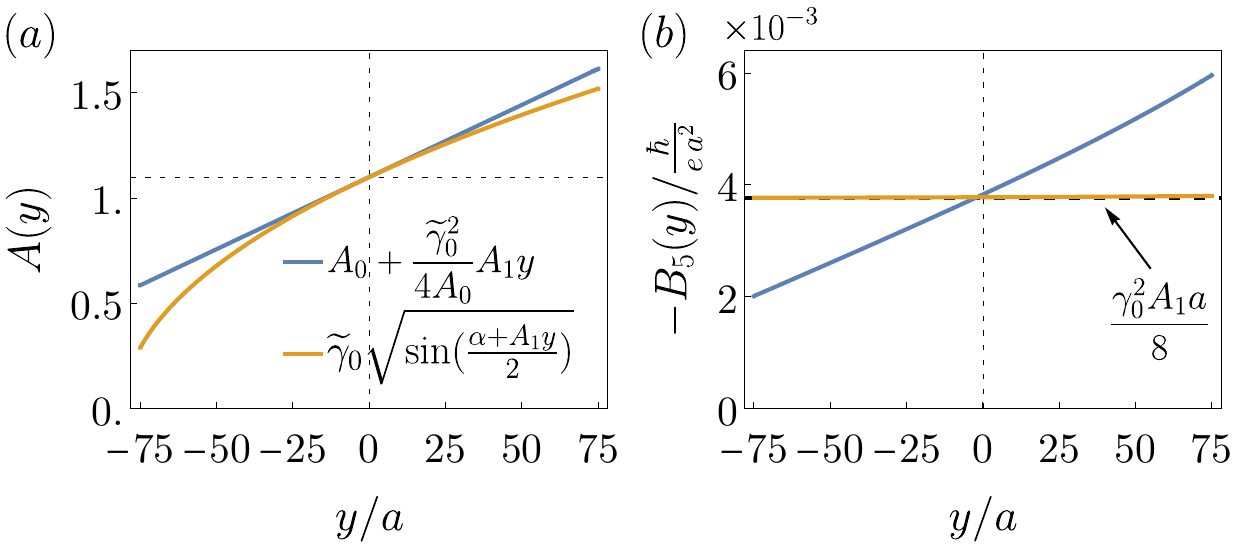}
    \caption{Different spatial profiles for the vector potential $A(y)$ and the corresponding pseudo-magnetic fields for $\G=1.1$, $A_1=0.004/a$, $N_y=150$.}
    \label{fig:AyB5}
\end{figure}

One can also consider other spatial profiles for the amplitude, for example,  $A(y)=A_0\left(1+\frac{\widetilde{\G}_0^2A_1}{4A_0^2}y\right)$, but the resulting $\vb{B}_5$ will always have a stronger spatial dependence as shown in Fig.~\ref{fig:AyB5}. We should point out that for the light considered in Eq.~\eqref{eq:linAy}, the spatial dependence is perpendicular to the propagation direction and we have neglected attenuation effects in the $x$-direction for simplicity. 

\begin{figure}[b]
    \centering
    \includegraphics[width=0.98\linewidth]{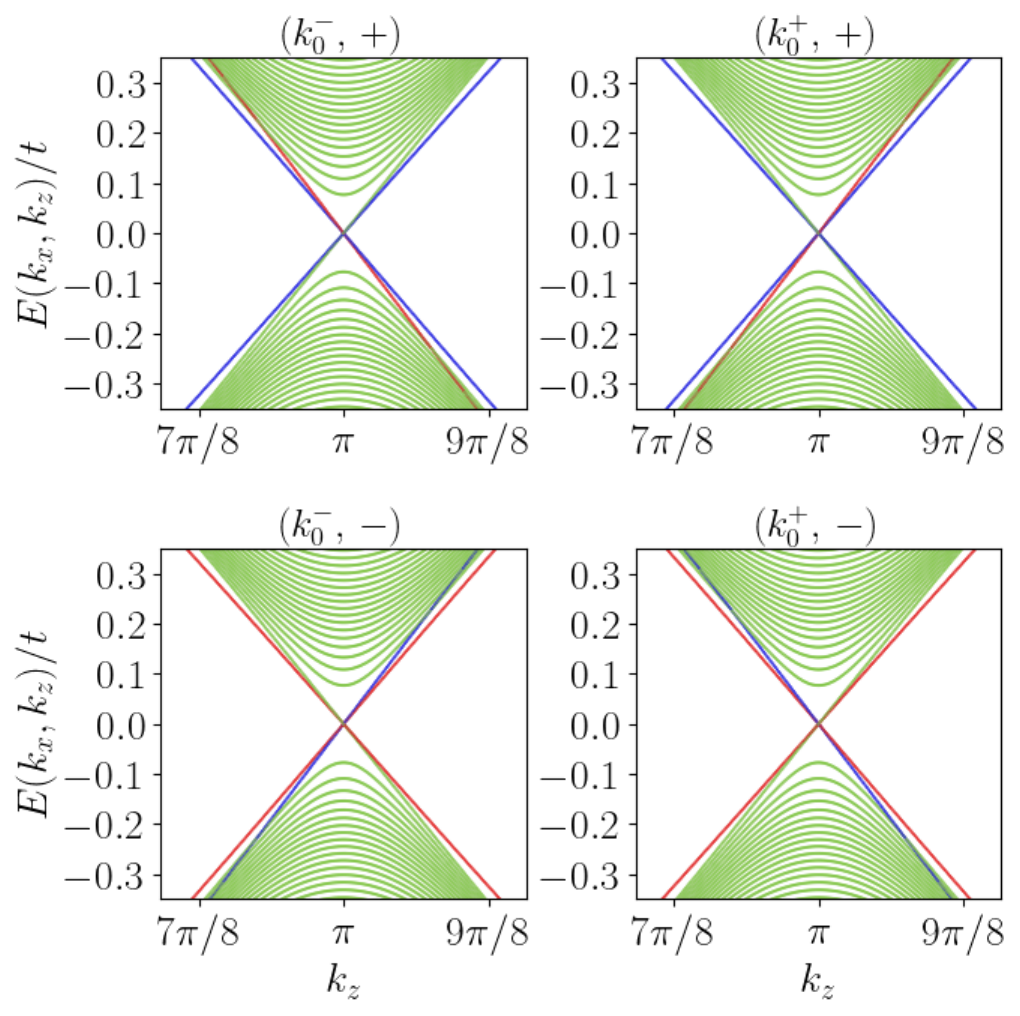}
    \caption{Energy spectrum around the separated R-nodes, $k_0^{-}$ and $k_0^{+}$, in presence of a pseudo-magnetic field generated from taking $A(y)$ from Eq.~\eqref{eq:Ay}. Green represents the bulk modes while red and blue represent the surface modes on opposite sides. The sgn($A_1$) is shown besides the node label. We have used $\G=1.1$, $|A_1|=0.004/a$, $N_y=150$.}
    \label{fig:llB5}
\end{figure}
With $A(y)$ in Eq.~\eqref{eq:Ay}, the resulting $4N_y\times 4N_y$ Hamiltonian is given in Appendix~\ref{appendix:HeffB5}.
The Landau level spectrum at the two nodes is given in Fig.~\ref{fig:llB5}. Since the nodes have the same chirality, the zeroth Landau levels disperse in the opposite direction.

\begin{figure}[t]
    \centering
    \includegraphics[width=0.7\linewidth]{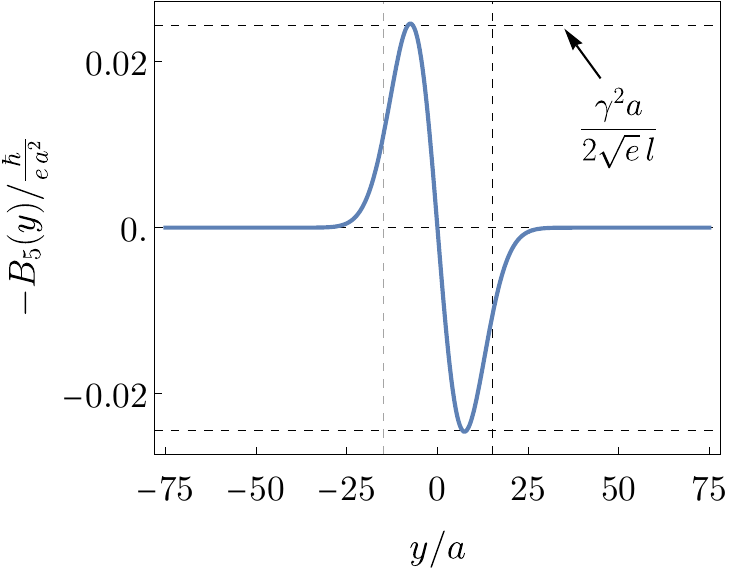}\hspace{1cm}
    \caption{Pseudo-magnetic domain wall for $A(y) = A_0 e^{-\frac{y^2}{l^2}}$. The vertical and horizontal dashed lines are at $\pm l/a$ and $\pm \frac{\G^2a}{2\sqrt{e}\,l}$, respectively. Here, $e$ is the base of the natural logarithm. We have used $\G=1.1$, $l=15a$, $N_y=150$.}
    \label{fig:B5_domain_wall}
\end{figure}

We also note that the spatial variation of light intensity in our Floquet approach naturally suggests the possibility of creating pseudo-magnetic domain walls. For example, one can consider a gaussian profile: 
\begin{align}
    A(y) &= A_0 e^{-\frac{y^2}{l^2}}, 
\end{align}
where $A_0<\widetilde{\G}_0$ and $l$ can be adjusted to tune the domain wall width. The $\vb{B}_5$ generated in this case is shown in Fig.~\ref{fig:B5_domain_wall}.
While real magnetic domain walls in Weyl semimetals have been studied both theoretically~\cite{araki2018localized, ozawa2024chiral, araki2020, araki2016universal, hannukainen2020axial}, and experimentally~\cite{He2023, fujiwara2024}, showing phenomena such as localized charge accumulation~\cite{araki2018localized}, and antisymmetric magnetoresistance~\cite{fujiwara2024}, light-induced pseudo-magnetic domain walls offer unique advantages. Unlike their magnetic counterparts which are pinned by material properties and defects, optically generated  pseudo-magnetic domain walls can be dynamically created, moved, and erased by adjusting the laser beam profile. Moreover, since $\vb{B}_5$ couples with opposite sign to different chiralities, these domain walls could exhibit physics absent in conventional magnetic systems, potentially including novel chiral transport phenomena and dynamically tunable topological states.

% ------------------------------------------------------------------------------------
\subsection{Real magnetic field}
\label{ssec:LL_B}

\begin{figure}[t]
    \centering
    \includegraphics[width=0.98\linewidth]{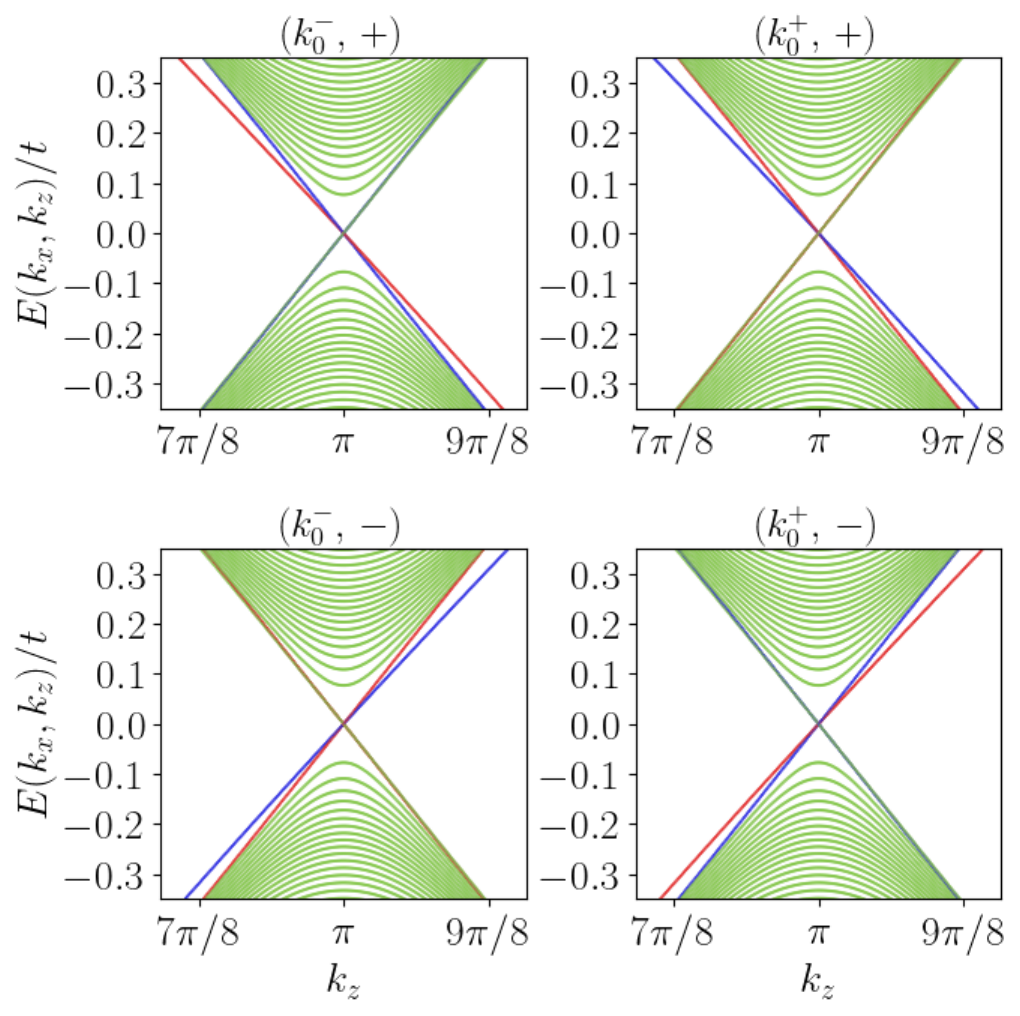}
    \caption{Energy spectrum around the R nodes, split using $A(y)=A_0$ for the drive amplitude, in presence of an external magnetic field described by $\vb{A}_{\rm ext}=(By,0,0)$ with $\frac{|B|a^2}{\hbar/e}=\frac{\G_0^2}{8}A_1a$. Green represents the bulk modes while red and blue represent the surface modes on opposite sides. sgn($B$) is shown besides the node label. We have used $\G=1.1$, $A_1=0.004/a$, $N_y=150$.}
    \label{fig:llB}
\end{figure}

\begin{figure}[t]
    \centering
    \includegraphics[width=0.98\linewidth]{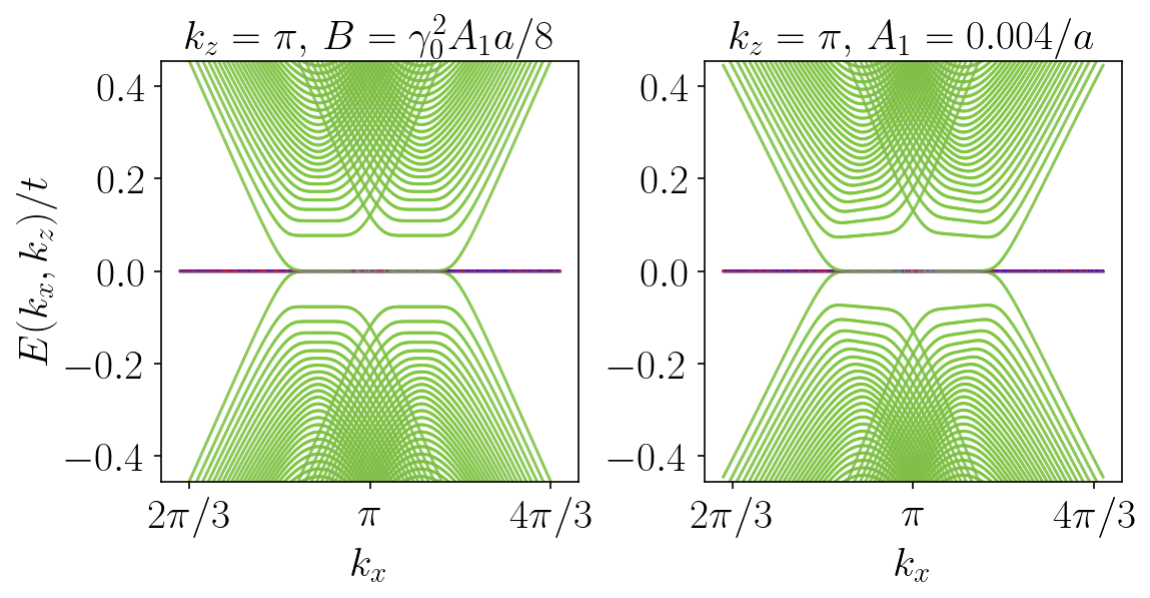}
    \caption{Energy spectrum along $k_x$ near the separated R-nodes in presence of an external magnetic field (left) and a pseudo-magnetic field (right). We have used $\G=1.1$, $N_y=150$.}
    \label{fig:ekx}
\end{figure}

For comparison, we now look at the case when the driven system from Eq.~\eqref{eq:driven0} is subject to a uniform real external magnetic field $\vb{B}=-B\vu{z}$ described by the vector potential $\vb{A}_{\rm ext}=(By,0,0)$. Since $\vb{A}_{\rm ext}$ has a $y$ dependence, $k_y$ is yet again not a good quantum number. The effect of the external magnetic field is included in the Hamiltonian via the Peierls substitution and the resulting $4N_y\times 4N_y$ Hamiltonian is given in Appendix~\ref{appendix:HeffB}. The Landau level spectrum at the two nodes is shown in Fig.~\ref{fig:llB}. We have chosen $B=\frac{\hbar}{ea^2} \frac{\G_0^2}{8} A_1a$ for a meaningful comparison with the $\vb{B}_5$ plots. We note that since the nodes have the same chirality, the zeroth Landau levels disperse in the same direction. However, this does not lead to a net equilibrium chiral magnetic effect, as the bulk chiral current is fully compensated by counter-propagating surface currents carried by the Fermi arcs~\cite{grushin2016inhomogeneous}.
We also compare the energy spectrum near the nodes for $\vb{B}$ and $\vb{B}_5$ fields along $k_x$ in Fig.~\ref{fig:ekx}.

In the following sections, we discuss the signatures of such $\vb{B}$ and $\vb{B}_5$ fields in both linear and nonlinear optical conductivity measurements.

% ------------------------------------------------------------------------------------
\section{Results for optical conductivity}
\label{sec:conductivity_results}

\subsection{First-order conductivity}
\label{sec:linear_cond}

\begin{figure}[t!]
    \centering
    \includegraphics[width=0.98\linewidth]{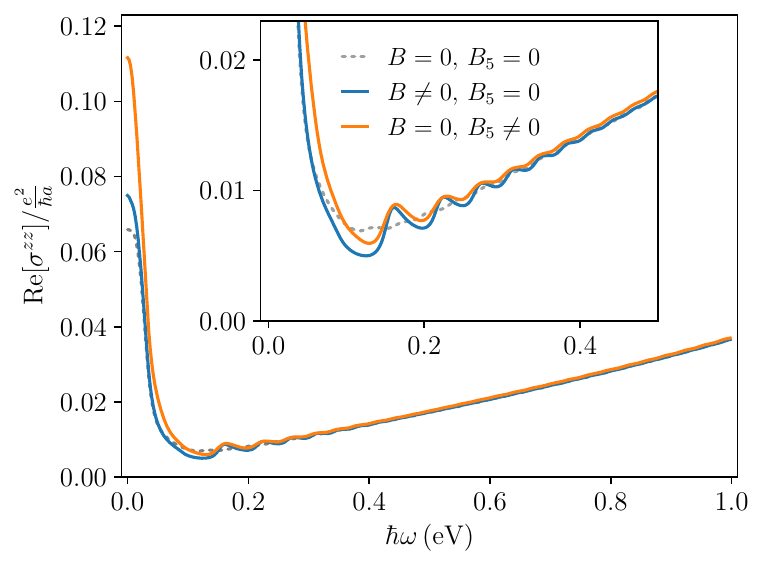}
    \caption{The main panel shows the real part of the longitudinal conductivity $\s^{zz}$ as a function of the probe frequency $\omega$ for $\vb{B}$ (blue), $\vb{B}_5$ (orange), and with neither of them (gray). In the inset, we zoom in to show oscillations for $\vb{B}$ and $\vb{B}_5$ fields arising from Landau level transitions. We have used $\G=1.1$, $N_y=150$, $\eta=0.01\,t$, $T=10$K.}
    \label{fig:loc_zz}
\end{figure}

\begin{figure}[t!]
    \centering
    \includegraphics[width=0.98\linewidth]{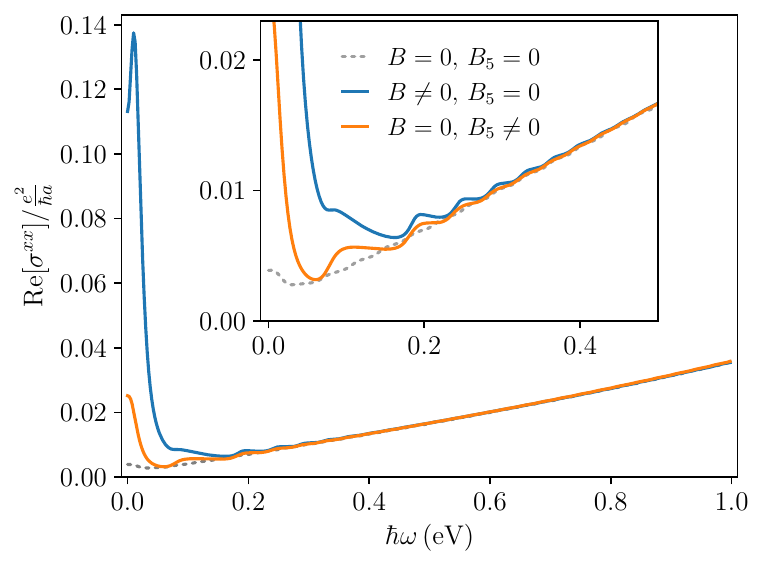}
    \caption{The main panel shows the real part of the longitudinal conductivity $\s^{xx}$ as a function of the probe frequency $\omega$ for $\vb{B}$ (blue), $\vb{B}_5$ (orange), and with neither of them (gray). In the inset, we zoom in to show oscillations for $\vb{B}$ and $\vb{B}_5$ fields arising from Landau level transitions. We have used $\G=1.1$, $N_y=150$, $\eta=0.01\,t$, $T=10$K.}
    \label{fig:loc_xx}
\end{figure}

With an effective Floquet Hamiltonian, we numerically calculate the longitudinal optical conductivities, specifically $\mathrm{Re}[\s^{zz}]$ and $\mathrm{Re}[\s^{xx}]$, for both real and pseudo-magnetic fields using~\cite{kumar2020linear},
\begin{align}
 \s^{bb}(\omega) &= \frac{e^2}{i\hbar V} \int\frac{\dd[2]{k}}{(2\pi)^2} \sum_{n\neq m} \frac{f_{nm}}{E_{nm}} \frac{|\bra{n}\hbar v^b\ket{m}|^2}{(\hbar\omega+i\eta-E_{nm})},
\end{align}
where $\omega$ is the probe frequency, $f_{nm}=f(E_n)-f(E_m)$ with $f(E_n)=1/(1+e^{E_n/k_BT})$, $E_{nm}=E_n-E_m$, $\hbar v^b=\pdv*{\hh^F}{k_b}$, and $V=N_ya$. We take $\eta=0.01\,t$ and $T=10\,\text{K}$. The results are shown in Figs.~\ref{fig:loc_zz} and \ref{fig:loc_xx}.

We see quantum oscillations in $\mathrm{Re}[\s^{zz}]$ and $\mathrm{Re}[\s^{xx}]$ for both real and pseudo-magnetic fields. These oscillations are a direct result of the underlying Landau level structure created by $\vb{B}$ and $\vb{B}_5$. This can be more clearly seen from the analytical expressions for these conductivities obtained using a low-energy linearized model as shown in Appendix~\ref{appendix:loc_theoretical}.

% ------------------------------------------------------------------------------------
\subsection{Second-order DC conductivity}
\label{sec:nonlinear_cond}

Having looked at the linear conductivity, we now calculate the injection conductivity for both real and pseudo-magnetic fields using
\begin{align}
\begin{split}
    \s_{\text{inj}}^{abc}(\omega) &= \frac{2\pi \tau e^3\hbar}{V} \int \frac{\dd[2]{k}}{(2\pi)^2} \sum_{n\neq m} f_{nm} (v_{nn}^a-v_{mm}^a) \\
    &\hspace{2.5cm} \times \frac{v_{nm}^b v_{mn}^c}{E_{nm}^2} \delta (E_{nm}-\hbar\omega),
\end{split}
\label{eq:injection}
\end{align}
where $\tau$ is the relaxation time that saturates the injection current. For all numerical calculations, we approximate the delta function with $\frac{1}{\pi}\frac{\eta}{(E_{nm}-\hbar\omega)^2+\eta^2}$. Since we have open boundary conditions in the $y$-direction, we calculate the corresponding velocity matrix using $\hbar v^y=\frac{1}{i}\qty[Y,\hh^F]$ where 
\begin{align}
    Y &= \mqty(\dmat{\tfrac{-N_y+1}{2}, \ddots, \tfrac{N_y-1}{2}}) \otimes \id_4 + \id_{N_y}\otimes \mqty(\dmat{0,\tfrac{1}{2},\tfrac{1}{2},0}) .
\end{align}

\begin{figure}[t!]
    \centering
    \includegraphics[width=0.98\linewidth]{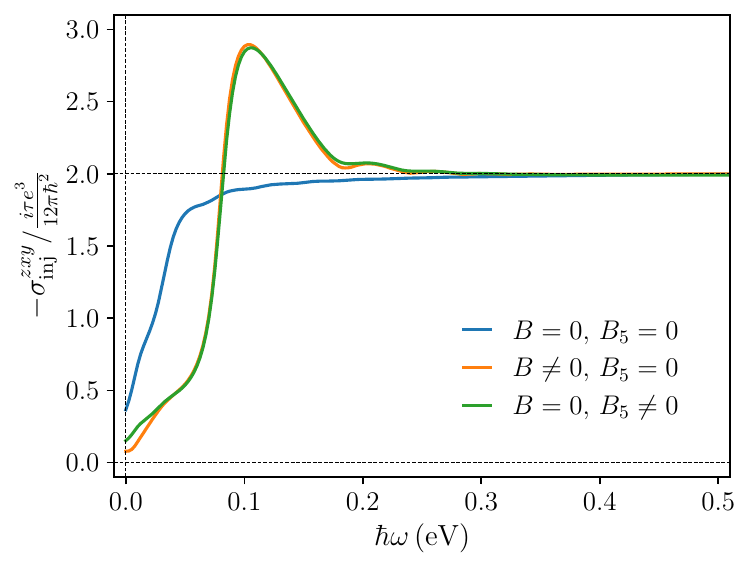}
    \caption{Injection conductivity $\s^{zxy}$ as a function of the probe frequency $\omega$ for $\vb{B}$ (orange), $\vb{B}_5$ (green), and both absent (blue). We have used $\G=1.1$, $A_1=0.004/a$, $N_y=150$, $\eta=0.01\,t$, $T=10$K.}
    \label{fig:soc_zxy}
\end{figure}

The results obtained for $\s_{\text{inj}}^{zxy}$ are shown in Fig.~\ref{fig:soc_zxy}. Since the system breaks inversion symmetry, a finite circular injection conductivity is allowed irrespective of whether $\vb{B}$ or $\vb{B}_5$ is present (the former does not break inversion symmetry but the latter does). As discussed earlier, the two nodes near the $\rm R$ point sit right at the chemical potential and have the same chirality which allow their contributions to add up. This results in the expected low-frequency plateau $\frac{12 \pi\hbar^2}{i\tau e^3} \s_{\text{inj}}^{zxy} = -2$ determined by their total chiral charge~\cite{raj2024photogalvanic}.
However, when a uniform $\vb{B}$ or $\vb{B}_5$ field is present, the response changes significantly and develops a low energy hump before settling to the zero field response value. This behavior is again a result of the Landau levels that these fields create (see Appendix~\ref{appendix:inj_theoretical}).

% ------------------------------------------------------------------------------------
\section{Conclusions}
\label{sec:conclusion}

In summary, we have shown that spatially varying linearly polarized light could provide a versatile route to engineer pseudo-magnetic fields in Weyl semimetals via Floquet band renormalization. Compared with strain engineering, the optical route is dynamically tunable and spatially selective, enabling on-demand control without material deformation. The model we studied has a charge-2 multifold Weyl point at $\rm R$ which could be split into two charge-1 Weyl points of the same chirality by shining linearly polarized light. We found that their separation depends on the light amplitude which can be given various spatial profiles to create desired $\vb{A}_5$ and $\vb{B}_5$, including the possibility of an optically generated dynamically tunable pseudo-magnetic domain wall.

In this study, we chose a specific light profile that resulted in a uniform pseudo-magnetic field. We computed the resulting Landau level spectrum and compared it to that of a real uniform magnetic field of the same strength. We found that the zeroth Landau level disperses in opposite directions for the former but in the same direction for the latter. We also examined the linear and injection conductivities for both $\vb{B}_5$ and $\vb{B}$ and found them to be qualitatively similar. We observed oscillations in the longitudinal linear optical conductivity coming from transitions between the lower Landau levels for both cases. For the injection conductivity, we found a low-energy hump followed by a plateau whose height was consistent with that of circular photogalvanic effect quantization. Together, these results provide a promising route for optically generating pseudo-gauge fields in topological semimetals and detecting them through optical conductivity measurements.

% ------------------------------------------------------------------------------------
\section{Acknowledgments}

We thank D. Sabsovich for helpful discussions. G.A.F. acknowledges funding from the National Science Foundation through DMR-2114825. G.A.F. acknowledges additional support from the Alexander
von Humboldt Foundation. S.C acknowledges support from JSPS KAKENHI (No. JP23H04865, No. 23K22487), MEXT, Japan.
 This work was supported by
the Israel Science Foundation (ISF) Grant No. 2307/24 (R.I.). R. I. is grateful for the hospitality of Perimeter Institute where part of this work was carried out. Research at Perimeter Institute is supported in part by the Government of Canada through the Department of Innovation, Science and Economic Development and by the Province of Ontario through the Ministry of
Colleges and Universities. This work was supported by a grant from the Simons Foundation (1034867,
Dittrich).

\appendix
%
% ------------------------------------------------------------------------------------
\section{Details of the Hamiltonian in Eq.~\eqref{eq:tb-static}}
\label{appendix:nn}

As pointed out in the supplemental material of Ref.~\cite{robredo2021new}, one needs to be careful to obtain the correct tight-binding model given in Eq.~\eqref{eq:tb-static}. We consider $s$-like spinless orbital at,
\begin{align}
\begin{split}
 r_A &= (\epsilon,\epsilon,\epsilon), \\
 r_B &= (\tfrac{1}{2}+\epsilon,\tfrac{1}{2}-\epsilon,-\epsilon), \\
 r_C &= (-\epsilon,\tfrac{1}{2}+\epsilon,\tfrac{1}{2}-\epsilon), \\
 r_D &= (\tfrac{1}{2}-\epsilon,-\epsilon,\tfrac{1}{2}+\epsilon), \\
\end{split}
\end{align}
and label the orbitals as A, B, C, and D, respectively. To get the correct lattice with symmetries of the space group \textit{P}$2_13$, we start with a small positive value for $\epsilon$ and determine all the nearest neighbors (six for each type, as shown in Table~\ref{tab:nn}), then set $\epsilon$ to $0$.
\begin{table}[th!]
\caption{Nearest neighbor coordinates}
\renewcommand{\arraystretch}{1.25}
\begin{ruledtabular}
\begin{tabular}{|l|l|l|}
& Nearest neighbors & Coordinate \\
\hline
\multirow{6}{*}{$\quad r_A$} & $r_B$            & $\tfrac{1}{2} (1,1,0)$   \\ 
                       & $r_B + (-1,0,0)$ & $\tfrac{1}{2} (-1,1,0)$  \\
                       & $r_C$            & $\tfrac{1}{2} (0,1,1)$   \\ 
                       & $r_C + (0,-1,0)$ & $\tfrac{1}{2} (0,-1,1)$  \\
                       & $r_D$            & $\tfrac{1}{2} (1,0,1)$   \\ 
                       & $r_D + (0,0,-1)$ & $\tfrac{1}{2} (1,0,-1)$  \\
\hline
\multirow{6}{*}{$\quad r_B$} & $r_A$            & $\tfrac{1}{2} (-1,-1,0)$ \\ 
                       & $r_A + (1,0,0)$  & $\tfrac{1}{2} (1,-1,0)$  \\
                       & $r_C + (1,0,0)$  & $\tfrac{1}{2} (1,0,1)$   \\ 
                       & $r_C + (1,0,-1)$ & $\tfrac{1}{2} (1,0,-1)$  \\
                       & $r_D + (0,0,-1)$ & $\tfrac{1}{2} (0,-1,-1)$ \\ 
                       & $r_D + (0,1,-1)$ & $\tfrac{1}{2} (0,1,-1)$  \\
\hline
\multirow{6}{*}{$\quad r_C$} & $r_A$            & $\tfrac{1}{2} (0,-1,-1)$ \\ 
                       & $r_A + (0,1,0)$  & $\tfrac{1}{2} (0,1,-1)$  \\
                       & $r_B + (-1,0,0)$ & $\tfrac{1}{2} (-1,0,-1)$ \\ 
                       & $r_B + (-1,0,1)$ & $\tfrac{1}{2} (-1,0,1)$  \\
                       & $r_D + (0,1,0)$  & $\tfrac{1}{2} (1,1,0)$   \\ 
                       & $r_D + (-1,1,0)$ & $\tfrac{1}{2} (-1,1,0)$  \\
\hline
\multirow{6}{*}{$\quad r_D$} & $r_A$            & $\tfrac{1}{2} (-1,0,-1)$ \\ 
                       & $r_A + (0,0,1)$  & $\tfrac{1}{2} (-1,0,1)$  \\
                       & $r_B + (0,0,1)$  & $\tfrac{1}{2} (0,1,1)$   \\ 
                       & $r_B + (0,-1,1)$ & $\tfrac{1}{2} (0,-1,1)$  \\
                       & $r_C + (0,-1,0)$ & $\tfrac{1}{2} (-1,-1,0)$ \\ 
                       & $r_C + (1,-1,0)$ & $\tfrac{1}{2} (1,-1,0)$  \\
\end{tabular}
\end{ruledtabular}
\renewcommand{\arraystretch}{1.0}
\label{tab:nn}
\end{table}

We should note that when taking open boundary condition along $y$, the nearest neighbors with a $+(.,\pm 1,.)$ will be in an adjacent unit cell.

\subsubsection*{Topological charge at the $\Gamma$ point}
\label{appendix:topo}

As discussed in the main text, near $\Gamma$,  the linearized Hamiltonian takes the following form
\begin{align}
\begin{split}
    \hh_{\Gamma}(\kv) &= 2t\big( \tau_0\s_x + \tau_x\s_x + \tau_x\s_0 \big) \\ 
    &\quad - t\big(k_x \tau_y\s_x + k_y \tau_0\s_y + k_z \tau_y\s_z \big),
\end{split}
\end{align}
which can be written as 
\begin{align}
\begin{split}
    \hh_{\Gamma}(\kv) &= 2t\big( \tau_0\s_z + \tau_z\s_z + \tau_z\s_0 \big) \\ 
    &\quad - t\big(k_x \tau_y\s_z + k_y \tau_0\s_y - k_z \tau_y\s_x \big),
\end{split}
\end{align}
after a $-\pi/2$ rotation of the $\sigma$ and $\tau$ matrices about the $y$ axis. This leads to  three-fold degenerate states with energy $-2t$ and a non-degenerate state with energy $6t$ at $k_x=k_y=k_z=0$. The corresponding eigenstates are given by:
$\ket{1}=\ket{\downarrow}\otimes\ket{\downarrow};\,\ket{2}=\ket{\uparrow}\otimes\ket{\downarrow};\,\ket{3}=\ket{\downarrow}\otimes\ket{\uparrow}$, and $\ket{4}=\ket{\uparrow}\otimes\ket{\uparrow}$, respectively. Here, we have used $\ket{\uparrow}=\smqty(1\\0)$ and $\ket{\downarrow}=\smqty(0\\1)$. When the Hamiltonian is projected in the basis $(\ket{3},-i\ket{2},\ket{1})$, it takes the following form:
\begin{align}
   \hh_{\Gamma}^{\rm proj}(\kv) &= t\begin{pmatrix}
       -2&& k_z&& ik_y\\k_z&&-2&& k_x\\-ik_y&&k_x&&-2
   \end{pmatrix}, \\
   &= -2t\,\mathbb{I}_3 + t(k_x\Lambda_1+k_y\Lambda_2+k_z\Lambda_3),
\end{align}
where $\Lambda_1$, $\Lambda_2$, $\Lambda_3$ are three Gell-Mann matrices
\begin{equation}
    \Lambda_1=\begin{pmatrix}
        0&&0&&0\\0&&0&&1\\0&&1&&0
    \end{pmatrix}; \Lambda_2=\begin{pmatrix}
        0&&0&&i\\0&&0&&0\\-i&&0&&0
    \end{pmatrix}; \Lambda_3=\begin{pmatrix}
        0&&1&&0\\1&&0&&0\\0&&0&&0
    \end{pmatrix}
\end{equation}
with $[\Lambda_i,\Lambda_j]=-i\epsilon_{ijk}\Lambda_k$ are closed under commutation and are isomorphic to $SU(2)$. This is equivalent to $S=1$ Weyl node which carries charge +2.

% ------------------------------------------------------------------------------------
\section{Effective Floquet Hamiltonian for circularly polarized light}
\label{appendix:heffCP}

We include the time periodic drive into our model from Eq.~\eqref{eq:tb-static} by substituting $\kv\to\kv+\frac{ea}{\hbar}\vb{A}(t)$ where the vector potential is as given by Eq.~\eqref{eq:circularA0}. Now, to obtain $\hh^F$, we first calculate $\hh^{(0)}= \frac{\Omega}{2\pi}\int_{0}^{2 \pi/\Omega} \dd{t} \hh(t)$ and $\hh^{(1)}=\frac{\Omega}{2\pi}\int_{0}^{2 \pi/\Omega} \dd{t} \hh(t) e^{-i \Omega t}$:
\begin{widetext}
\begin{gather}
\scalemath{0.9}{%
\hh^{(0)} = 2t \mqty(
 0 &
 e^{\tfrac{i}{2}k_y}\cos\!\left(\tfrac{k_x}{2}\right) J_0\!\left(\tfrac{\gamma}{2}\right) &
 e^{\tfrac{i}{2}k_z}\cos\!\left(\tfrac{k_y}{2}\right) J_0\!\left(\tfrac{\gamma}{\sqrt{2}}\right) &
 e^{\tfrac{i}{2}k_x}\cos\!\left(\tfrac{k_z}{2}\right) J_0\!\left(\tfrac{\gamma}{2}\right) \\
 e^{-\tfrac{i}{2}k_y}\cos\!\left(\tfrac{k_x}{2}\right) J_0\!\left(\tfrac{\gamma}{2}\right) &
 0 &
 e^{\tfrac{i}{2}k_x}\cos\!\left(\tfrac{k_z}{2}\right) J_0\!\left(\tfrac{\gamma}{2}\right) &
 e^{-\tfrac{i}{2}k_z}\cos\!\left(\tfrac{k_y}{2}\right) J_0\!\left(\tfrac{\gamma}{\sqrt{2}}\right) \\
 e^{-\tfrac{i}{2}k_z}\cos\!\left(\tfrac{k_y}{2}\right) J_0\!\left(\tfrac{\gamma}{\sqrt{2}}\right) &
 e^{-\tfrac{i}{2}k_x}\cos\!\left(\tfrac{k_z}{2}\right) J_0\!\left(\tfrac{\gamma}{2}\right) &
 0 &
 e^{\tfrac{i}{2}k_y}\cos\!\left(\tfrac{k_x}{2}\right) J_0\!\left(\tfrac{\gamma}{2}\right) \\
 e^{-\tfrac{i}{2}k_x}\cos\!\left(\tfrac{k_z}{2}\right) J_0\!\left(\tfrac{\gamma}{2}\right) &
 e^{\tfrac{i}{2}k_z}\cos\!\left(\tfrac{k_y}{2}\right) J_0\!\left(\tfrac{\gamma}{\sqrt{2}}\right) &
 e^{-\tfrac{i}{2}k_y}\cos\!\left(\tfrac{k_x}{2}\right) J_0\!\left(\tfrac{\gamma}{2}\right) &
 0),
}
\end{gather}
\begin{gather}
\scalemath{0.9}{%
\hh^{(1)} = 2t \mqty(
0 
& e^{\tfrac{i}{2}k_y}\cos\!\Big(\tfrac{k_x}{2}\Big)\,J_1\!\Big(\tfrac{\gamma}{2}\Big) 
& i\,e^{\tfrac{i}{2}k_z}\sin\!\Big(\tfrac{k_y}{2}+\tfrac{\pi}{4}\Big)\,J_1\!\Big(\tfrac{\gamma}{\sqrt{2}}\Big) 
& -\,e^{\tfrac{i}{2}k_x}\sin\!\Big(\tfrac{k_z}{2}\Big)\,J_1\!\Big(\tfrac{\gamma}{2}\Big) \\
-\,e^{-\tfrac{i}{2}k_y}\cos\!\Big(\tfrac{k_x}{2}\Big)\,J_1\!\Big(\tfrac{\gamma}{2}\Big) 
& 0 
& -\,e^{\tfrac{i}{2}k_x}\sin\!\Big(\tfrac{k_z}{2}\Big)\,J_1\!\Big(\tfrac{\gamma}{2}\Big) 
& i\,e^{-\tfrac{i}{2}k_z}\sin\!\Big(\tfrac{k_y}{2}-\tfrac{\pi}{4}\Big)\,J_1\!\Big(\tfrac{\gamma}{\sqrt{2}}\Big) \\
i\,e^{-\tfrac{i}{2}k_z}\sin\!\Big(\tfrac{k_y}{2}-\tfrac{\pi}{4}\Big)\,J_1\!\Big(\tfrac{\gamma}{\sqrt{2}}\Big) 
& -\,e^{-\tfrac{i}{2}k_x}\sin\!\Big(\tfrac{k_z}{2}\Big)\,J_1\!\Big(\tfrac{\gamma}{2}\Big) 
& 0 
& e^{\tfrac{i}{2}k_y}\cos\!\Big(\tfrac{k_x}{2}\Big)\,J_1\!\Big(\tfrac{\gamma}{2}\Big) \\
-\,e^{-\tfrac{i}{2}k_x}\sin\!\Big(\tfrac{k_z}{2}\Big)\,J_1\!\Big(\tfrac{\gamma}{2}\Big) 
& i\,e^{\tfrac{i}{2}k_z}\sin\!\Big(\tfrac{k_y}{2}+\tfrac{\pi}{4}\Big)\,J_1\!\Big(\tfrac{\gamma}{\sqrt{2}}\Big) 
& -\,e^{-\tfrac{i}{2}k_y}\cos\!\Big(\tfrac{k_x}{2}\Big)\,J_1\!\Big(\tfrac{\gamma}{2}\Big) 
& 0),
}
\end{gather}
\end{widetext}
where, $J_1(x)$ is the first-order Bessel function of the first kind, and $\G=\frac{A_0a}{\hbar/e}$. We found the Jacobi–Anger expansions quite useful in evaluating these integrals. Since $\hh^{(-1)}={\hh^{(1)}}^{\dagger}$, we get  
\begin{widetext}
\begin{gather}
\scalemath{0.9}{%
\frac{\Big[\hh^{(1)},\hh^{(-1)}\Big]}{\hbar\Omega} = \frac{4t^2}{\hbar\Omega} J_1\!\left(\tfrac{\gamma}{\sqrt{2}}\right)^2 \mqty(
 \sin{k_y} &
 -2 \sqrt{2}\,\xi \, e^{\tfrac{i}{2}k_z} \Xi_{s_3} &
 -4 \xi^2 e^{\tfrac{i}{2}k_x}
   \Xi_{c_2s} &
 -2 i \sqrt{2}\,\xi \, e^{\tfrac{i}{2}k_y}
   \Xi_{c_3} \\
 -2 \sqrt{2}\,\xi \, e^{-\tfrac{i}{2}k_z}
   \Xi_{s_3} &
 -\sin{k_y} &
 -2 i \sqrt{2}\,\xi \, e^{-\tfrac{i}{2}k_y}
   \Xi_{c_3} &
 4 \xi^2 e^{\tfrac{i}{2}k_x}
   \Xi_{c_2s} \\
 -4 \xi^2 e^{-\tfrac{i}{2}k_x}
   \Xi_{c_2s} &
 2 i \sqrt{2}\,\xi \, e^{\tfrac{i}{2}k_y}
   \Xi_{c_3} &
 -\sin{k_y} &
 2 \sqrt{2}\,\xi \, e^{-\tfrac{i}{2}k_z}
   \Xi_{s_3} \\
 2 i \sqrt{2}\,\xi \, e^{-\tfrac{i}{2}k_y} \Xi_{c_3} &
 4 \xi^2 e^{-\tfrac{i}{2}k_x}
   \Xi_{c_2s} &
 2 \sqrt{2}\,\xi \, e^{\tfrac{i}{2}k_z}
   \Xi_{s_3} &
 \sin{k_y}),
}
\end{gather}
\end{widetext}
where we have defined $\xi = J_1\!\left(\tfrac{\gamma}{2}\right)/J_1\!\left(\tfrac{\gamma}{\sqrt{2}}\right)$ and 
\begin{align}
\Xi_{s_3}=&\sin\!\left(\tfrac{k_x}{2}\right)\sin\!\left(\tfrac{k_y}{2}\right)\sin\!\left(\tfrac{k_z}{2}\right),\\
\Xi_{c_3}=&\cos\!\left(\tfrac{k_x}{2}\right)\cos\!\left(\tfrac{k_y}{2}\right)\cos\!\left(\tfrac{k_z}{2}\right),\\
\Xi_{c_2s}=&\cos\!\left(\tfrac{k_x}{2}\right)\cos\!\left(\tfrac{k_y}{2}\right)\sin\!\left(\tfrac{k_z}{2}\right).
\end{align}

% ------------------------------------------------------------------------------------
\section{Deriving Eq.~\eqref{eq:efh_LR_lin}}
\label{appendix:transformation}

The linearized version of Eq.~\eqref{eq:efh_LR} reads
\begin{align}
\begin{split}
\widetilde{H}^{F,L}_R(\kv) &= t \Big( k_x \tau_0 \s_y + k_y \tau_y\s_z  + k_z \tau_y \s_x + \tfrac{\G^2}{4} \tau_x\s_0 \Big).
\end{split}
\end{align}
First, we rotate the  $\tau$ matrices about the $x$-axis by $\pi/2$ which replaces $\tau_y \rightarrow \tau_z$ to give
\begin{align}
\begin{split}
\widetilde{H}^{F,L}_R(\kv) &= t \Big( k_x \tau_0 \s_y + k_y \tau_z \s_z  + k_z \tau_z \s_x + \tfrac{\G^2}{4} \tau_x \s_0 \Big).
\end{split}
\end{align}
Next, we take the unitary matrix, $U$, given by:
\begin{equation}
    U=\frac{(\tau_0+\tau_z)}{2}\s_0+\frac{(\tau_0-\tau_z)}{2}\s_y,
\end{equation}
which satisfies
\begin{align}
    U \, (\tau_0\s_y) \, U^\dagger &= \tau_0\s_y, \\
    U \, (\tau_z\s_{x,z}) \, U^\dagger &= \tau_0\s_{x,z}, \\
    U \, (\tau_x\s_0) \, U^\dagger &= \tau_x\s_y.
\end{align}
Applying this unitary transformation gives 
\begin{align}
\widetilde{H}^{F,L}_R(\kv) &= t \Big( k_x\tau_0\s_y + k_y\tau_0\s_z + k_z\tau_0\s_x + \tfrac{\G^2}{4} \tau_x\s_y \Big).
\end{align}
Finally, we rotate both sets of Pauli matrices about the $y$-axis by $-\pi/2$ which results in 
\begin{align} 
\hh^{F,L}_R(\kv) &= t\left(   k_x\tau_0 \s_y - k_y \tau_0\s_x + k_z \tau_0\s_z  + \tfrac{\G^2}{4}\tau_z \s_y\right),
\end{align}
which is Eq.~\eqref{eq:efh_LR_lin}.

% ------------------------------------------------------------------------------------
\section{Effective Floquet Hamiltonian for linearly polarized light with open boundary condition along the $y$-axis}
\label{appendix:Heff}

% ------------------------------------------------------------------------------------
\subsection{$\vb{B}_5$ along $\vu{z}$}
\label{appendix:HeffB5}

We consider the linearly polarized light described by Eqs.~\eqref{eq:linAy} and \eqref{eq:Ay}. For a term like $t\,c^{\dagger}_{\vb{r}_j}c_{\vb{r}_i}$, the vector potential modifies the hopping amplitude $t$ via the Peierls substitution:
\begin{align}
 t_{i\rightarrow j} &= t\,e^{-i\frac{\sin(\Omega t)}{\hbar/e}\int_{\vb{r}_i}^{\vb{r}_j} A(y)(\dd{y}+\dd{z})}, \\
                    &= t\,e^{-i\frac{\sin(\Omega t)}{\hbar/e} (y_j+z_j-y_i-z_i) \int_0^1 A(y_i+\rho(y_j-y_i))\dd{\rho}},
\end{align}
where we used $\vb{r}_i=(x_i,y_i,z_i)$ and $\vb{r}_j=(x_j,y_j,z_j)$. We compute this for hopping between all nearest neighbor sites given in Table~\ref{tab:nn} and then Fourier transform $x$, $z$ to $k_x$, $k_z$ to go from $c_{\vb{r}_i}\to c_{y_i}(k_x,k_z)$. This gives us a $4N_y\times 4N_y$ time-dependent Hamiltonian $H(t)$. Finally, we get the effective Floquet Hamiltonian as $\hh^F = \hh^{(0)} = \frac{\Omega}{2\pi}\int_{0}^{2 \pi/\Omega} \dd{t} H(t)$ since the $1/\Omega$ term is zero. The $4\times 4$ blocks that make it are given by
\vspace{0.5cm}
\begin{widetext}
\begin{gather}
\scalemath{0.9}{
\hh_{m,m} = t\mqty(
 0 & 2\cos \left(\frac{k_x}{2}\right) J_0\left(\frac{\G_0\,\mathcal{E}_2(m,m+\frac{1}{2}) }{2}\right) & e^{\frac{i}{2} k_z} J_0\left(\G_0\,\mathcal{E}_2\left(m,m+\frac{1}{2}\right) \right) & 2 e^{\frac{i}{2} k_x} \cos \left(\frac{k_z}{2}\right) J_0\left(\frac{\G_0\, \mathcal{E}(m) }{2}\right) \\
 2 \cos \left(\frac{k_x}{2}\right) J_0\left(\frac{\G_0\,\mathcal{E}_2(m,m+\frac{1}{2})}{2}\right) & 0 & 2 e^{\frac{i}{2} k_x} \cos \left(\frac{k_z}{2}\right) J_0\left(\frac{\G_0\, \mathcal{E}(m+\frac{1}{2}) }{2}\right) & e^{-\frac{i}{2}k_z} J_0\left(\G_0\,\mathcal{E}_2\left(m,m+\frac{1}{2}\right)\right) \\
 e^{-\frac{i}{2}k_z} J_0\left(\G_0\,\mathcal{E}_2\left(m,m+\frac{1}{2}\right)\right) & 2 e^{-\frac{i}{2} k_x} \cos \left(\frac{k_z}{2}\right) J_0\left(\frac{\G_0\, \mathcal{E}(m+\frac{1}{2}) }{2}\right) & 0 & 0 \\
 2 e^{-\frac{i}{2}k_x} \cos \left(\frac{k_z}{2}\right) J_0\left(\frac{\G_0\,\mathcal{E}(m)}{2}\right) & e^{\frac{i}{2}k_z} J_0\left(\G_0\,\mathcal{E}_2\left(m,m+\frac{1}{2}\right)\right) & 0 & 0 )}, \\[10pt]
\scalemath{0.9}{
\hh_{m+1,m} = t\mqty(
 0 & 0 & e^{\frac{i}{2}k_z} & 0 \\
 0 & 0 & 0 & 0 \\
 0 & 0 & 0 & 0 \\
 0 & e^{\frac{i}{2}k_z} & 2 \cos \left(\frac{k_x}{2}\right) J_0\left(\frac{\G_0\,\mathcal{E}_2(m+\frac{1}{2},m+1)}{2}\right) & 0 ),
\qquad\quad \hh_{m-1,m} = t\mqty(
 0 & 0 & 0 & 0 \\
 0 & 0 & 0 & e^{-\frac{i}{2}k_z} \\
 e^{-\frac{i}{2}k_z} & 0 & 0 & 2 \cos \left(\frac{k_x}{2}\right) J_0\left(\frac{\G_0\,\mathcal{E}_2(m-\frac{1}{2},m)}{2}\right) \\
 0 & 0 & 0 & 0 )},
\end{gather}
\end{widetext}
where $m$ corresponds to the $y$ coordinate and runs from $\frac{-N_y+1}{2},\ldots,\frac{N_y-1}{2}$. Additionally, we have defined
\begin{align}
\mathcal{E}(m) &= \sqrt{\sin(\tfrac{\alpha+aA_1 m}{2})}, \\
\mathcal{E}_2(m,n) &= \tfrac{8}{A_1}  \Big[ E\left(\left. \tfrac{\pi-(\alpha+aA_1m)}{4} \right|2\right) - E\left(\left. \tfrac{\pi-(\alpha+aA_1n)}{4} \right|2\right) \Big],
\end{align}
where $E(x|2)$ is the incomplete Elliptic integral of the second kind which is real valued over $-\frac{\pi}{4}\leq x \leq \frac{\pi}{4}$. For faster numerical evaluation of $E(x|2)$, we used the Carlson’s symmetric forms. Most importantly, this choice of $A(y)$ results in $\vb{B}_5\approx -\frac{\hbar}{ea} \frac{\G_0^2}{8} A_1\vu{z}$ which has almost no spatial dependence.

% ------------------------------------------------------------------------------------
\subsection{$\vb{B}$ along $\vu{z}$}
\label{appendix:HeffB}

As before, we consider the linearly polarized light described by Eq.~\eqref{eq:linAy} but with $A(y)=A_0$. This implies that $\vb{B}_5=0$ and instead we consider a uniform real magnetic field $\vb{B}=-B\vu{z}$ described by the vector potential $\vb{A}_{\rm ext}=(By,0,0)$. Thus, we have 
\begin{align}
    \vb{A}_{\rm total} = \left( By,A_0\sin(\Omega t),A_0\sin(\Omega t) \right),
\end{align}
which changes the hopping amplitude to
\begin{align}
 t_{i\rightarrow j} &= t\,e^{-i\frac{B}{\hbar/e}\int_{\vb{r}_i}^{\vb{r}_j} y\dd{x}} e^{-i\frac{A_0\sin(\Omega t)}{\hbar/e}\int_{\vb{r}_i}^{\vb{r}_j}(\dd{y}+\dd{z})}, \\
                    &= t\,e^{-i\frac{B}{\hbar/e} (x_j-x_i)\tfrac{(y_j+y_i)}{2}} e^{-i\frac{A_0\sin(\Omega t)}{\hbar/e} (y_j+z_j-y_i-z_i)}.
\end{align}
Following the same steps as in Appendix~\ref{appendix:HeffB5}, we get the effective Floquet Hamiltonian with 
\begin{widetext}
\begin{gather}
\scalemath{0.9}{
\hh_{m,m} = t\mqty(
 0 & 2\cos \left(\frac{k_x+\Bt(m+\frac{1}{4})}{2}\right) J_0\left(\frac{\G}{2}\right) & e^{\frac{i}{2} k_z} J_0(\G) & 2 e^{\frac{i}{2} (k_x+\Bt m)} \cos \left(\frac{k_z}{2}\right) J_0\left(\frac{\G}{2}\right) \\
 2 \cos \left(\frac{k_x+\Bt(m+\frac{1}{4})}{2}\right) J_0\left(\frac{\G}{2}\right) & 0 & 2 e^{\frac{i}{2} (k_x+\Bt(m+\frac{1}{2}))} \cos \left(\frac{k_z}{2}\right) J_0\left(\frac{\G}{2}\right) & e^{-\frac{i}{2}k_z} J_0(\G) \\
 e^{-\frac{i}{2}k_z} J_0(\G) & 2 e^{-\frac{i}{2} (k_x+\Bt(m+\frac{1}{2}))} \cos \left(\frac{k_z}{2}\right) J_0\left(\frac{\G}{2}\right) & 0 & 0 \\
 2 e^{-\frac{i}{2}(k_x+\Bt m)} \cos \left(\frac{k_z}{2}\right) J_0\left(\frac{\G}{2}\right) & e^{\frac{i}{2}k_z} J_0(\G) & 0 & 0 )}, \\
\scalemath{0.9}{
 \hh_{m+1,m} = t\mqty(
 0 & 0 & e^{\frac{i}{2}k_z} & 0 \\
 0 & 0 & 0 & 0 \\
 0 & 0 & 0 & 0 \\
 0 & e^{\frac{i}{2}k_z} & 2 \cos \left(\frac{k_x+\Bt(m+\frac{3}{4})}{2}\right) J_0\left(\frac{\G}{2}\right) & 0 ),
\qquad\quad \hh_{m-1,m} = t\mqty(
 0 & 0 & 0 & 0 \\
 0 & 0 & 0 & e^{-\frac{i}{2}k_z} \\
 e^{-\frac{i}{2}k_z} & 0 & 0 & 2 \cos \left(\frac{k_x+\Bt(m-\frac{1}{4})}{2}\right) J_0\left(\frac{\G}{2}\right) \\
 0 & 0 & 0 & 0 )},
\end{gather}
\end{widetext}
where we have defined $\G=\frac{A_0a}{\hbar/e}$ and $\Bt=\frac{Ba^2}{\hbar/e}$. As before, $m$ corresponds to the $y$ coordinate and runs from $\frac{-N_y+1}{2},\ldots,\frac{N_y-1}{2}$.

% ------------------------------------------------------------------------------------
\section{Landau levels and optical conductivities}

We consider a Weyl node with finite tilt in the $z$-direction and described by
\begin{align}
    H &= \hbar\mqty( u_zk_z + u_tk_z - \frac{\mu}{\hbar} & u(k_x-ik_y)\\ 
    u(k_x+ik_y) & -u_zk_z + u_tk_z - \frac{\mu}{\hbar}).
\end{align}
The chirality of the node is given by $\chi=\textrm{sgn}(u_z)$. We will assume $u>0$. Throughout this section, $k_x$, $k_y$, $k_z$ have the units of inverse length.

A uniform magnetic field, $\vb{B}=B\vu{z}$, is applied to the system (we will assume $B>0$). In Landau gauge this is effected by the vector potential $\vb{A}=(-By,0,0)$ which couples by making $\kv\to \kv+\frac{e}{\hbar}\vb{A}$. Note that if we were to do this analysis for $\vb{A}_5=(-By,0,0)$ which produces $\vb{B}_5=B\vu{z}$, the only difference would be in the way the pseudo-vector potential couples to the system: $\kv\to \kv+\chi\frac{e}{\hbar}\vb{A}_5$. This implies that for $\chi=+1$ all the results we obtain hereafter for $\vb{B}$ will directly apply whereas for $\chi=-1$ the results for $B<0$ will apply.
Since $\vb{A}$ has a $y$ dependence, $k_y$ is no longer a good quantum number. Presence of the magnetic field changes the Hamiltonian to
\begin{align}
    H &= \hbar\mqty( u_zk_z + u_tk_z - \frac{\mu}{\hbar} & u(k_x-\frac{eB}{\hbar}y-\partial_y)\\ 
    u(k_x-\frac{eB}{\hbar}y+\partial_y) & -u_zk_z + u_tk_z - \frac{\mu}{\hbar}).
\end{align}
We now define
\begin{align}
    b &=  \frac{l_B}{\sqrt{2}}\left(\frac{y-y_0}{l_B^2}+\partial_y\right), \\
    b^\dagger &=  \frac{l_B}{\sqrt{2}}\left(\frac{y-y_0}{l_B^2}-\partial_y\right),
\end{align}
where $l_B=\sqrt{\frac{\hbar}{eB}}$ is the magnetic length and $y_0=k_xl_B^2$. These operators satisfy $[b,b^\dagger]=1$. Since $b,b^\dagger$ behave as the lowering and raising operators, respectively, we can define 
\begin{align}
    b\ket{n} &= \sqrt{n}\ket{n-1} \\
    b^\dagger\ket{n} &= \sqrt{n+1}\ket{n+1}
\end{align}
where $\ket{n}$ are the eigenstates of the number operator, $b^\dagger b\ket{n}=n\ket{n}$. The eigenstates are easily found to be $\ket{n}=\frac{(b^\dagger)^n}{\sqrt{n!}}\ket{0}=\frac{2^{-n/2}}{\sqrt{n!}}H_n\left(\frac{y-y_0}{l_B}\right)\ket{0}$ where $H_n(x)$ are the Hermite polynomials and $\ket{0}=\frac{1}{\sqrt[4]{\pi l_B^2}}e^{-\frac{(y-y_0)^2}{2l_B^2}}$ is obtained by requiring $b\ket{0}=0$.

In terms of these operators, the Hamiltonian can be rewritten as
\begin{align}
    H &= \hbar\mqty( u_zk_z + u_tk_z - \frac{\mu}{\hbar} & -\frac{\sqrt{2}u}{l_B}b \\ 
    -\frac{\sqrt{2}u}{l_B}b^\dagger & -u_zk_z + u_tk_z - \frac{\mu}{\hbar}).
\end{align}
It is easy to see that $\psi=\mqty(\A\ket{n-1}\\ \B\ket{n})$ can be eigenstate of $H$ for a suitably chosen $\A,\B$ (with $\A^2+\B^2=1$). Solving for $H\psi=E\psi$ we get the following system of coupled linear equations 
\begin{align}
    \left(u_zk_z + u_tk_z - \frac{\mu}{\hbar} - \frac{E}{\hbar}\right)\A -\frac{\sqrt{2n}u}{l_B}\B &= 0, \\
    -\frac{\sqrt{2n}u}{l_B}\A + \left(-u_zk_z + u_tk_z - \frac{\mu}{\hbar} - \frac{E}{\hbar}\right)\B &= 0.
\end{align}
Here, we need to consider two cases:
\begin{enumerate}
    \item $n=0$: We are forced to set $\A=0$ which results in the eigenstate $\psi_0=\mqty(0\\ \ket{0})$ with the eigenenergy
    \begin{align}
        E_0(k_z) = -\hbar(u_z-u_t)k_z -\mu.
    \end{align}
    \item $n\neq 0$: Solving for the eigenenergy gives
    \begin{align}
        E_n^{\pm}(k_z) = -\mu + \hbar u_tk_z \pm\hbar|u_z| \sqrt{k_z^2+\frac{2n}{l_B^2}\frac{u^2}{u_z^2}}.
    \end{align}
    The corresponding eigenstates are given by $\psi_n^{\pm}=\mqty(\A_n^{\pm}\ket{n-1}\\ \B_n^{\pm}\ket{n})$ with 
    \begin{align}
        \A_n^{\pm} &= \tfrac{\frac{\sqrt{2n}u}{l_B}}{\sqrt{\left(\frac{\sqrt{2n}u}{l_B}\right)^2+\left(u_zk_z\mp \sqrt{u_z^2k_z^2+2n\frac{u^2}{l_B^2}}\,\right)^2}}, \\
        \B_n^{\pm} &= \tfrac{u_zk_z\mp \sqrt{u_z^2k_z^2+2n\frac{u^2}{l_B^2}}}{\sqrt{\left(\frac{\sqrt{2n}u}{l_B}\right)^2+\left(u_zk_z\mp \sqrt{u_z^2k_z^2+2n\frac{u^2}{l_B^2}}\,\right)^2}}.
    \end{align}
\end{enumerate}

% ------------------------------------------------------------------------------------
\subsection{Calculating $\s^{abc}_{\rm inj}$}
\label{appendix:inj_theoretical}

The injection conductivity is given by
\begin{align}
    \s_{\text{inj}}^{abc} &= \frac{\tau e^3\hbar}{ l_B^2}\int \frac{\dd{k_z}}{2\pi} \sum_{s\neq s'} f_{ss'}\frac{(v_{ss}^a-v_{s's'}^a) v_{ss'}^b v_{s's}^c}{E_{ss'}^2}  \delta (E_{ss'}-\hbar\omega),
\end{align}
where $s,s'$ are the Landau level indices with $\ket{s}=\ket{n,\lambda}$ and $\lambda=\{+,-\}$.
We need to find the velocity matrix elements first. The velocity operators are given by
\begin{align}
    \hat{v}_x &= u\s_x, \\
    \hat{v}_y &= u\s_y ,\\
    \hat{v}_z &= u_t\mathbb{I}_2 + u_z\s_z. 
\end{align}
For matrix elements involving only the nonzero LLs, we have
\begin{align}
\begin{split}
    v^x_{ss'} &= u\mel{n,\lambda}{\s_x}{n',\lambda'} \\
              &= u(\A_{n}^{\lambda}\B_{n'}^{\lambda'}\delta_{n-1,n'} + \B_{n}^{\lambda}\A_{n'}^{\lambda'}\delta_{n+1,n'}),
\end{split}\\
\begin{split}
    v^y_{ss'} &= u\mel{n,\lambda}{\s_y}{n',\lambda'} \\
              &= -iu(\A_{n}^{\lambda}\B_{n'}^{\lambda'}\delta_{n-1,n'} -  \B_{n}^{\lambda}\A_{n'}^{\lambda'}\delta_{n+1,n'}),
\end{split}\\
\begin{split}
    v^z_{ss'} &= \mel{n,\lambda}{u_t\mathbb{I}_2+u_z\s_z}{n',\lambda'} \\
              &= \left((u_t+u_z)\A_{n}^{\lambda}\A_{n'}^{\lambda'}+(u_t-u_z)\B_{n}^{\lambda}\B_{n'}^{\lambda'}\right)\delta_{n,n'}.
\end{split}
\end{align}
For matrix elements involving the zeroth LL, we have
\begin{align}
    v^x_{00} &= v^y_{00} = v^z_{s0} = 0, \\
    v^z_{00} &= u_t-u_z, \\
    v^x_{s0} &= u\A_1^{\lambda}\delta_{n,1}, \\
    v^y_{s0} &= -iu\A_1^{\lambda}\delta_{n,1},
\end{align}
where $\ket{s}\neq \psi_0$.
Given these velocity matrix elements, it is easy to see that $\s^{xyz}_{\rm inj}=\s^{yzx}_{\rm inj}=0$. The only nonzero CPGE tensor is given by (assuming $\omega>0$, $u_t=0$, $\mu=0$)
\begin{widetext}
\begin{align}
    \s^{zxy}_{\rm inj} &= -i\frac{\tau e^3}{\hbar^2}\frac{\textrm{sgn}(u_z)}{2\pi} \frac{\zeta^2}{1+\zeta^2} \left( 1 + \frac{1}{1-\zeta^2}\sum_{n=1}^{\left\lfloor \frac{1}{4}\left(\frac{1}{\zeta}-\zeta\right)^2 \right\rfloor} \sqrt{(1-\zeta^2)^2-4n\,\zeta^2} \right) \Theta\left(1-\zeta^2\right), \label{eq:zxy_analytical}
\end{align}
\end{widetext}
where we have defined $\zeta=\frac{\sqrt{2}u}{\omega l_B}$ and 
\begin{equation}
   \Theta(x) =
    \begin{cases}
      0 & \text{if $x\leq 0$} \\
      1 & \text{if $x>0$}
    \end{cases}.
\end{equation}

% ------------------------------------------------------------------------------------
\subsection{Calculating $\s^{zz}$}
\label{appendix:loc_theoretical}

The longitudinal linear conductivity is given by
\begin{align}
 \s^{zz}(\omega) &= \frac{-ie^2\hbar}{2\pi l_B^2} \int\frac{\dd{k_z}}{2\pi} \sum_{s\neq s'} \frac{f_{ss'}}{E_{ss'}} \frac{|v^z_{ss'}|^2}{(\hbar\omega+i\eta-E_{ss'})}.
\end{align}
Let us focus on its real part, given by
\begin{align}
 \textrm{Re}[\s^{zz}(\omega)] &= \frac{-e^2\hbar}{2l_B^2} \int\frac{\dd{k_z}}{2\pi} \sum_{s\neq s'} f_{ss'}\frac{|v^z_{ss'}|^2}{E_{ss'}} \delta(E_{ss'}-\hbar\omega).
\end{align}
With $\mu=0$, $u_t=0$ and $\omega>0$, we get
\begin{align}
\begin{split}
 \textrm{Re}[\s^{zz}(\omega)] &= \frac{e^2\hbar u_z^2}{4\pi l_B^2} \sum_{n\geq 1} \int \dd{k_z} \frac{(\A_{n}^{+}\A_{n}^{-}-\B_{n}^{+}\B_{n}^{-})^2}{E_n^+-E_n^-} \hspace{1cm}\\
 &\hspace{2.5cm} \times\delta(E_n^+-E_n^- - \hbar\omega).
\end{split}
\end{align}
The delta function constraint gives
\begin{align}
    k_z = \pm \sqrt{\frac{\omega^2}{4u_z^2}-\frac{2n}{l_B^2}\frac{u^2}{u_z^2}},
\end{align}
which would be real if 
\begin{align}
    n\leq \left\lfloor\frac{\omega^2 l_B^2}{8u^2}\right\rfloor.
\end{align}
\begin{figure*}[t]
    \centering
    \subfloat[Real part of the linear conductivity $\s^{zz}$]{
    \includegraphics[width=0.33\linewidth]{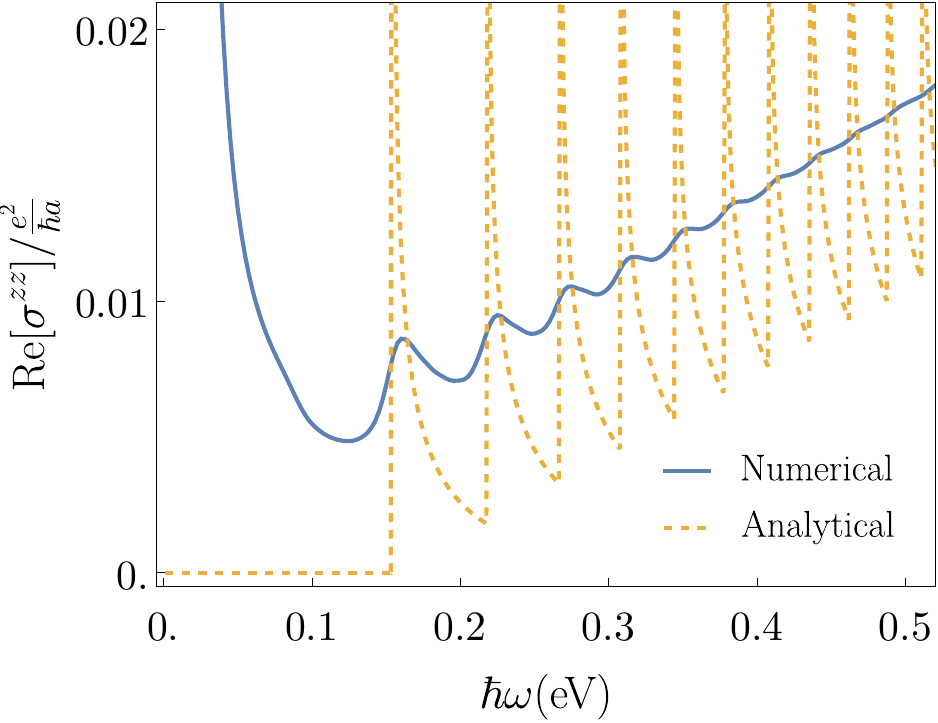}}
    \hfill
    \subfloat[Real part of the linear conductivity $\s^{xx}$]{
    \includegraphics[width=0.33\linewidth]{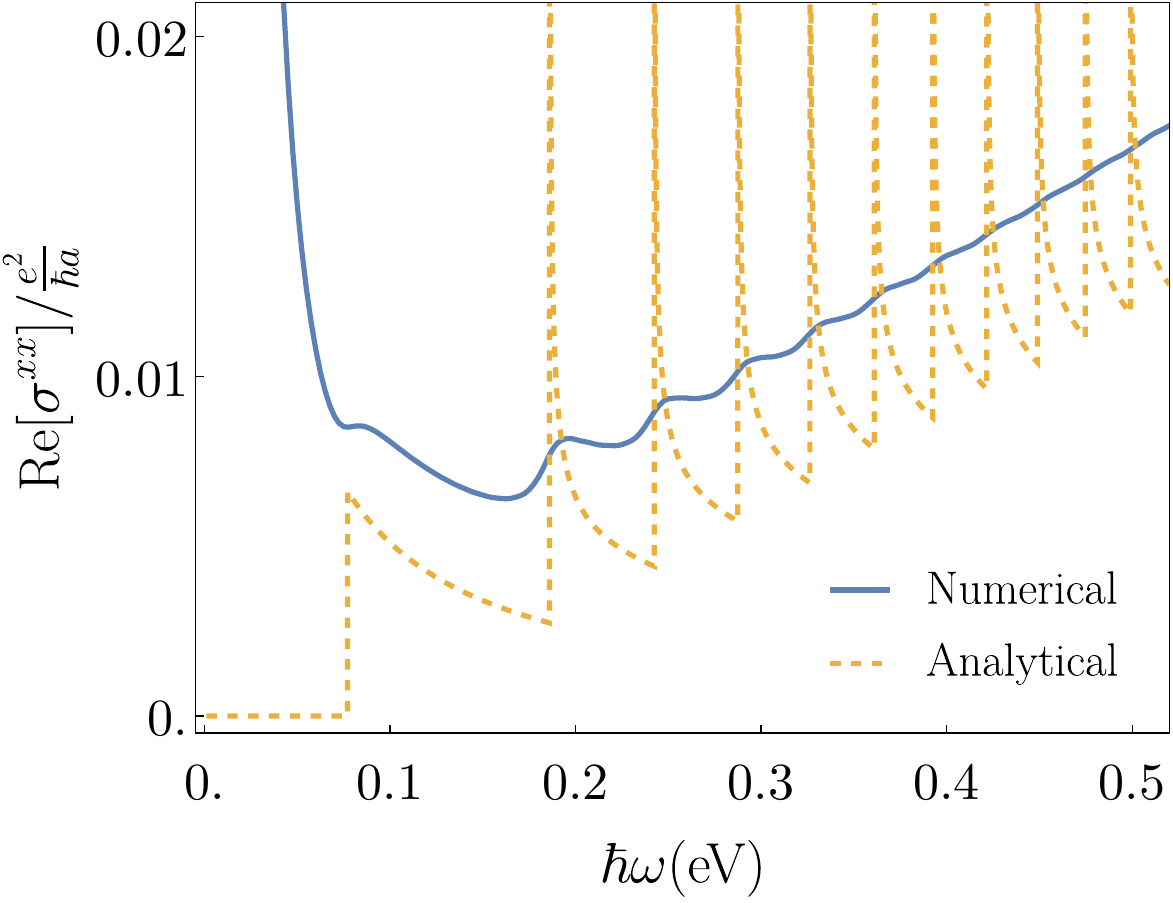}}
    \hfill
    \subfloat[Injection conductivity $\s^{zxy}$]{
    \includegraphics[width=0.31\linewidth]{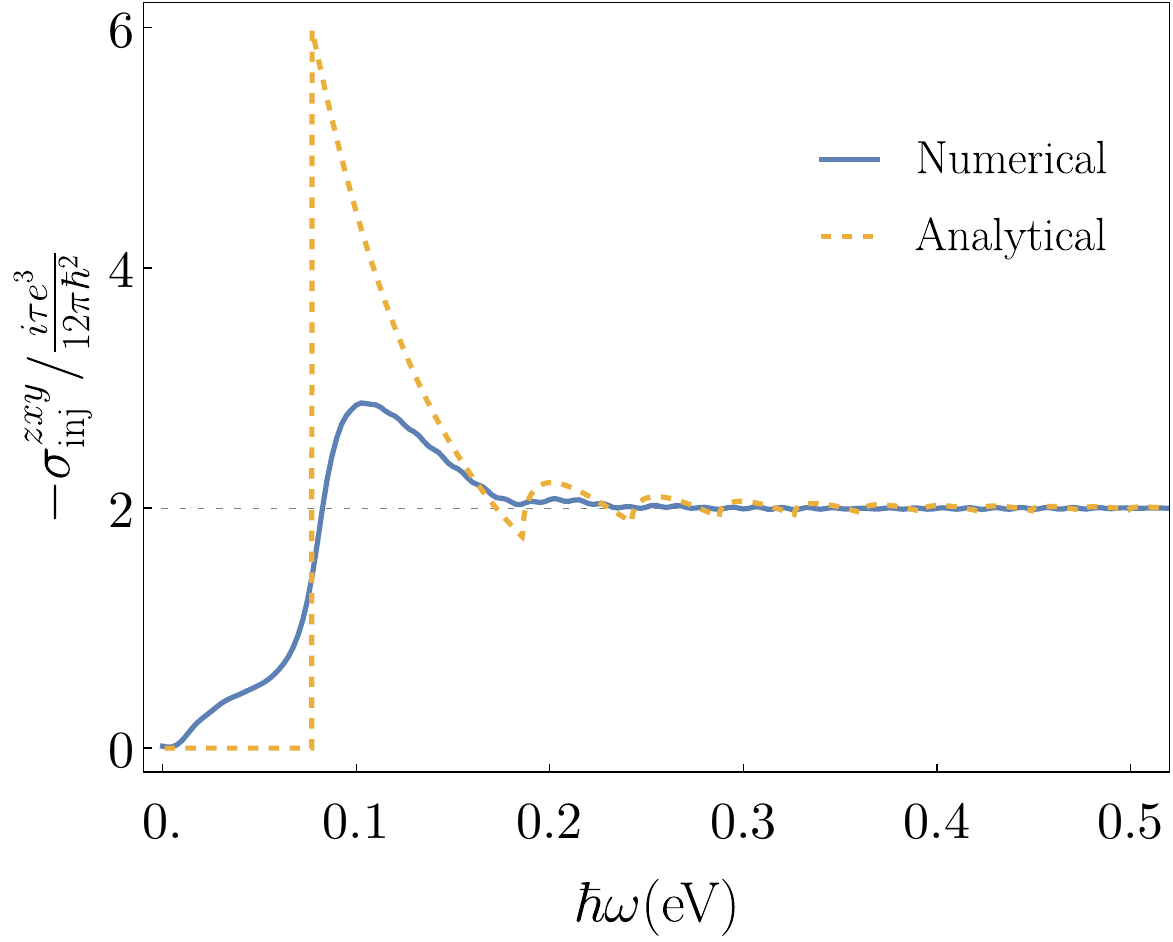}}
    \caption{The numerically computed conductivities for uniform magnetic field from Figs.~\ref{fig:loc_xx}, \ref{fig:loc_zz} and \ref{fig:soc_zxy} are shown in blue. We compare them against the analytical results (orange) applied to its low energy linearized version.}
    \label{fig:compare_th_num}
\end{figure*}
Thus, we have
\begin{widetext}
\begin{align}
 \textrm{Re}[\s^{zz}(\omega)] &= \frac{e^2}{\hbar a} \frac{|u_z|}{u}\frac{a}{l_B}\frac{\zeta^2}{2\sqrt{2} \pi } \sum_{n=1}^{\left\lfloor\frac{1}{4\zeta^2}\right\rfloor} \frac{n}{\sqrt{\frac{1}{4\zeta^2}-n}} \, \Theta\left(\frac{1}{4\zeta^2}-1\right). \label{eq:zz_analytical}
\end{align}
\end{widetext}

Similarly, we calculate the real part of $\sigma^{xx}$ to obtain 
\begin{widetext}
\begin{align}
 \textrm{Re}[\s^{xx}(\omega)] &= \frac{e^2}{\hbar a} \frac{u}{|u_z|}\frac{a}{l_B}\frac{\zeta}{4\sqrt{2} \pi } \left( 1 + \sum_{n=1}^{\left\lfloor \frac{1}{4}\left(\frac{1}{\zeta}-\zeta\right)^2 \right\rfloor} \frac{1-(1+2n)\zeta^2}{\sqrt{\left(1-\zeta^2\right)^2-4n\,\zeta^2}}\right)\,\Theta\left(1-\zeta^2\right). 
 \label{eq:xx_analytical}
\end{align}
\end{widetext}%

We can now compare the numerically obtained full tight binding model results from earlier against this. An important point to note here is that we need to replace $u$ with $\sqrt{u_xu_y}$ before using Eqs.~\eqref{eq:zxy_analytical}, \eqref{eq:zz_analytical} and \eqref{eq:xx_analytical} as the slope of the Weyl cone is not the same along $k_x$ and $k_y$ when $A_0$ is finite. In fact, for our model in Eq.~\eqref{eq:driven0}, we have
\begin{align}
    u_x &= \frac{ta}{\hbar} J_0\left(\frac{\G}{2}\right)\sqrt{1-\left(\tfrac{1-J_0(\G)}{2J_0(\G/2)}\right)^2}, \\
    u_y &= \frac{ta}{\hbar} \frac{1+J_0(\G)}{2}, \\
    u_z &= \frac{ta}{\hbar} J_0\left(\frac{\G}{2}\right).
\end{align}
Additionally, we need to multiply the analytical result by $2$ to account for both nodes. The comparison is shown in Fig.~\ref{fig:compare_th_num}. We find that the theoretical results align well with the numerical results.

\bibliography{Refs_updated.bib}
\end{document}